\documentclass[aps,twocolumn,floatfix,superscriptaddress]{revtex4-1}

\usepackage{graphicx}
\usepackage{indentfirst}
\usepackage{latexsym}
\usepackage{multirow}
\usepackage{tabls}
\usepackage{epsfig}
\usepackage{multirow}
\usepackage{subfigure}
\usepackage{comment}
\usepackage{hyperref}

\usepackage{color}
\usepackage[usenames,dvipsnames]{xcolor}
\usepackage{amssymb}

\usepackage{amsfonts}
\usepackage{amsmath}
\usepackage{bm}
\usepackage{mathrsfs}

\usepackage{algorithm}
\usepackage{algpseudocode} 
\usepackage{setspace} 

\newcommand{\mb}[1]{\mbox{\boldmath $#1$}}
\newcommand{\cI}{{\cal I}}
\newcommand{\be}{\begin{equation}}
\newcommand{\ee}{\end{equation}}

\newcommand{\mlam}{{\bm{\lambda}}}
\newcommand{\mLam}{{\bm{\Lambda}}}

\def\lsim{\mathrel{\rlap{\lower3pt\hbox{\hskip1pt$\sim$}}
    \raise1pt\hbox{$<$}}}                
\def\gsim{\mathrel{\rlap{\lower3pt\hbox{\hskip1pt$\sim$}}
    \raise1pt\hbox{$>$}}}         

\def\coordeq{ \, \mathrel{ \rlap{\hbox{\hskip-2.5pt$=$} }
    \raise4pt\hbox{$\cdot$}} \, }                

\begin{document}

\title{Fast prediction and evaluation of gravitational waveforms using surrogate models}

\def\addUMDa{Center for Scientific Computation and Mathematical Modeling, University of Maryland, College Park, MD 20742, USA}

\def\addUMDb{Department of Physics, Joint Space Sciences Institute,  Maryland Center for Fundamental Physics, University of Maryland, College Park, MD 20742, USA}

\def\addCaltech{Theoretical Astrophysics, California Institute of Technology,
Pasadena, CA, 91125, USA}

\def\addBrown{Division of Applied Mathematics, Brown University, Providence, RI 02912, USA}

\def\addEPFL{EPFL-SB-MATHICSE, \'{E}cole Polytechnique F\'{e}d\'{e}rale de Lausanne (EPFL), CH-1015 Lausanne, Switzerland}

\author{Scott E. Field}
\affiliation{\addUMDb}

\author{Chad R. Galley}
\affiliation{\addCaltech}

\author{Jan S. Hesthaven}
\affiliation{\addEPFL}

\author{Jason Kaye}
\affiliation{\addBrown}

\author{Manuel Tiglio}
\affiliation{\addUMDa}
\affiliation{\addUMDb}
\affiliation{\addCaltech}

\begin{abstract}
We propose a solution to the problem of quickly and accurately predicting gravitational waveforms within any given physical model. The method is relevant for both real-time applications and in more traditional scenarios where the generation of waveforms using standard methods can be prohibitively expensive. Our approach is based on three offline steps resulting in an accurate reduced-order model in both parameter and physical dimensions that can be used as a surrogate for the true/fiducial waveform family.
First, a set of $m$ parameter values is determined using a greedy algorithm from which a reduced basis representation is constructed. Second, these $m$ parameters induce the selection of $m$ time values for interpolating a waveform time series using an empirical interpolant that is built for the fiducial waveform family. 
Third, a fit in the parameter dimension is performed for the waveform's value at each of these $m$ times. The cost of predicting $L$ waveform time samples for a generic parameter choice is of order ${\cal O} \left( m L + m c_{\rm fit} \right)$ online operations where $c_{\rm fit}$ denotes the fitting function operation count and, typically, $m \ll L$. The result is a compact, computationally efficient, and accurate surrogate model that retains the original physics of the fiducial waveform family while also being fast to evaluate. We generate accurate surrogate models for Effective One Body (EOB) waveforms of non-spinning binary black hole coalescences with durations as long as $10^5 M$, mass ratios from $1$ to $10$, and for multiple spherical harmonic modes. We find that these surrogates are more than three orders of magnitude faster to evaluate as compared to the cost of generating EOB waveforms in standard ways. Surrogate model building for other waveform families and models follow the same steps and have the same low computational online scaling cost. For expensive numerical simulations of binary black hole coalescences we thus anticipate extremely large speedups in generating new waveforms with a surrogate. As waveform generation is one of the dominant costs in parameter estimation algorithms and parameter space exploration, surrogate models offer a new and practical way to dramatically accelerate such studies without impacting accuracy.
\end{abstract}

\maketitle

\section{Introduction}

A direct detection of gravitational waves generated by the coalescence of a compact binary system is among the most anticipated discoveries to be made in gravitational wave physics. The signal from such an event will codify perhaps the only attainable information about the existence, dynamics, and underlying physics of the strongest gravitating objects in the universe. Currently, there are few, if any, direct observations pertaining to gravity in the strong field regime but there is enough data to show agreement with the predictions of general relativity when gravitational fields and speeds are not too large \cite{Will:2005va,Stairs:2003eg}. 

In the case of binary black holes, where the fields and speeds can be large, one must rely on numerical simulations of the Einstein equations to discover how these systems evolve. The resulting solution depends on the choice of initial data. The intrinsic parameter space of binary black holes in quasi-circular orbit is seven dimensional, consisting of the mass ratio and the three spin angular momentum components for each black hole. Different choices of parameters can lead to qualitatively different outcomes, such as the final speed of the merged black hole due to a ``kick'' from the asymmetric emission of gravitational waves \cite{2007PhRvL..99d1102K,Pollney:2007ss,Gonzalez:2006md,Gonzalez:2007hi,Herrmann:2007ac,Herrmann:2007zz,Herrmann:2007ex,Campanelli:2007cga, Campanellietal:ApJ659, Campanelli:2007ew,Lousto:2008dn,Rezzolla:2010df}. In addition, potentially interesting effects due to strong precession from highly spinning black holes are waiting to be discovered and understood. Unfortunately, each numerical relativity (NR) simulation typically involves the use of large scale supercomputers, making an exploration of the parameter space a currently computationally intractable problem. For example, one might employ a uniform or random sampling strategy of the parameter space that, for a mere $4$ points per dimension, requires $4^7 \approx 16,\!000$ expensive numerical solutions of binary black hole coalescences. This number, while still being a very coarse survey of the parameter space, is substantially greater (by more than an order of magnitude) than all the simulations performed by all of the numerical relativity groups to date  \cite{Mroue:2013xna,Pekowsky:2013ska,Ajith:2012az,Hinder:2013oqa}. 

To help alleviate this computational bottleneck, models of the inspiral, merger, and ringdown phases of a binary black hole (BBH) coalescence have been developed over the last decade \cite{Buonanno:1998gg,Ajith:2007qp,Damour:2009ic,Buonanno:2009qa,Pan:2009wj,Santamaria:2010yb,Sturani:2010yv,Pan:2011gk,Abadie:2011kd,Damour:2012ky,Taracchini:2012ig}. The purpose of these phenomenological models is to provide a sufficiently accurate representation of a BBH waveform within some range of parameters by fitting certain coefficients and functions to a set of waveforms extracted from numerical simulations. In doing so, the models help to reduce the amount of information needed to represent NR waveforms. While these models are significantly faster than solving the Einstein field equations they remain computational bottlenecks for parameter estimation studies, which typically require generating millions of waveforms on the fly. Additionally, they still rely on waveforms computed from numerical simulations of binary black hole mergers and are thus unable, at least currently, to accurately model gravitational waveforms throughout the entire seven-dimensional parameter space, although efforts to attack this problem are underway \cite{Hinder:2013oqa,Pan:2013rra}. 

Other important considerations come from precessing inspirals of compact binaries such as binary neutron stars. 
Generating the corresponding waveforms requires solving a set of ordinary differential equations (ODEs) and substituting the solutions into the post-Newtonian expressions for the phase and amplitude corrections. 
Given that around $520,\!000$ to $860,\!000$ waveforms are needed to build template banks for just non-precessing, slowly spinning binary neutron stars for advanced LIGO \cite{Brown:2012qf}, which would already be a computational challenge, it follows that the large number of ODE solves would be prohibitively expensive in the general precessing case. Waveform generation for precessing compact binary inspirals constitutes the main computational bottleneck for both template bank construction and parameter estimation studies. 

In this paper we offer a solution to the need for cheap and accurate generation of gravitational waveforms, that may otherwise be too expensive to compute for the application of interest. Alternative waveform prediction methods have recently been proposed~\cite{Cannon:2012gq,Smith:2012du,PhysRevD.85.081504,brown_sc_2013_13} (see Appendix~\ref{sec:donot} for a brief discussion and Ref.~\cite{brown_sc_2013_13} for comparison details, especially). These works have focused on gravitational waveform models known through closed-form expressions while the focus of this paper is on those described by differential equations. To achieve this, it is crucial to take advantage of the rich structure underlying the waveforms of interest. Importantly, our method builds accurate surrogate models that do not sacrifice the underlying physics but instead combines the efficiency and power of reduced order modeling techniques with high accuracy sparse representations and an offline-online decomposition of the problem. 

Work over the last few years has shown that gravitational waveforms exhibit redundancy in the parameter space \cite{Galley:2010rc, Cannon:2010qh, Field:2011mf, Cannon:2011xk, Caudill:2011kv, PhysRevD.86.084046}, suggesting that the amount of information necessary to represent a fiducial waveform model is smaller than might be anticipated.  This reduction can be captured accurately using only a remarkably few number $m$ of representative waveforms. These $m$ representative waveforms can be found using a greedy algorithm and comprise a \emph{reduced basis} \cite{Field:2011mf} from which all other waveforms within the same physical model can be represented provided one can compute their projections onto the basis. In practice, this is neither feasible nor worthwhile because projecting onto the basis requires already knowing the waveform that one is seeking to represent in the first place. This is particularly the case for waveform families that are expensive to generate, such as those from numerical relativity (NR) simulations of the full Einstein equations. Instead, we aim to use \emph{only} the information provided by the $m$ representative waveforms of the reduced basis to {\em predict} waveforms accurately and cheaply for any desired parameter values. 

To accomplish this goal we first build the reduced basis as mentioned above and described in more detail in Sec.~\ref{sec:Greedy} and Appendix \ref{sec:RBapp}.
Second, we construct a temporal interpolant \footnote{The steps taken are exactly the same for waveforms in the frequency domain.} whereby any fiducial waveform is fully specified through its evaluation at $m$ appropriately chosen times. While this may seem surprising it is important to recall that there are only $m$ independent pieces of information in the waveform family as indicated by the $m$ waveforms that comprise the reduced basis. Indeed, we will show that the $m$ reduced basis uniquely specify these $m$ specially chosen times. The interpolation method outlined above, which is called \emph{empirical interpolation} because it generates an interpolant specific to the given fiducial waveform family, takes advantage of this nearly optimal representation strategy in parameters to provide a corresponding representation strategy in time. See Sec.~\ref{sec:EIM} and Appendix \ref{sec:EIMapp} for more details. Finally, at each empirical interpolation time we perform a fit in the parameter dimension of the waveform's amplitude and phase. Evaluating these fits yield $m$ time samples from which the waveform is accurately recovered through its empirical interpolant representation. Remarkably, the outlined method allows for a waveform within any physical model to be predicted for any parameter value of interest based solely on a knowledge of $m$ fiducial waveforms. 

Combining these pieces of information yields a \emph{surrogate model} for the fiducial waveform family. The method to build the surrogate has several useful properties. First, the method is entirely hierarchical, i.e. the accuracy of the surrogate model can be improved, if necessary, by \emph{adding} fiducial waveforms without discarding any of the previous ones. Second, the surrogate model can be evaluated using only ${\cal O}\left( m L + m c_{\rm fit} \right)$ computational operations, where $L$ is the number of time samples at which the model is evaluated and $c_{\rm fit}$ is the typical fitting function operation count. This provides a significant speedup compared to the usual way that fiducial waveforms are generated, as we demonstrate below with a surrogate model for non-spinning Effective One Body (EOB) waveforms. The speedup compared to numerical simulations of the full Einstein equations is expected to be significantly larger. 

\section{Surrogate waveform models} \label{sec:approach}

We denote the gravitational waveform produced from a fiducial model by $h(t;\mlam)$. Here, $t$ denotes time and $\mlam$ is the waveform parameterization (e.g., mass ratio and spins). We denote the surrogate model of the fiducial waveform family by $h_{\rm S}(t;\mlam)$ and describe its construction in this section. 

When numerically generating waveforms, by solving partial or ordinary differential equations, one typically solves an initial (or initial-boundary) value problem  
for a fixed $\mlam_i$ thereby generating ${h(t;\mlam_i)}$ on a densely sampled grid in time. In this paper we develop a procedure for building $h_{\rm S}(t;\mlam)$ through judicious choices of $\mlam_i$ and the corresponding output ${h(t;\mlam_i)}$ found by solving the relevant equations defining the fiducial problem. Crucially, given the complexity of existing numerical solvers, our approach to surrogate modeling is intentionally non-intrusive to legacy codes. 

We seek a {\em minimal} number of $\mlam_i$ selections for a target accuracy such that the surrogate has a comparable or smaller error than that associated with the underlying waveform model. This is important both for the speed of evaluating the surrogate model and for overcoming computational challenges with building it in cases where one cannot generate $h(t;\mlam_i)$ for arbitrarily many values of $\mlam$. Naturally, if more data is available it should be possible to include it and improve the surrogate's quality. This means that the surrogate model should be hierarchical by construction, improving as more simulations become available and without discarding previous ones.

\begin{figure}
\includegraphics[width=0.98\linewidth]{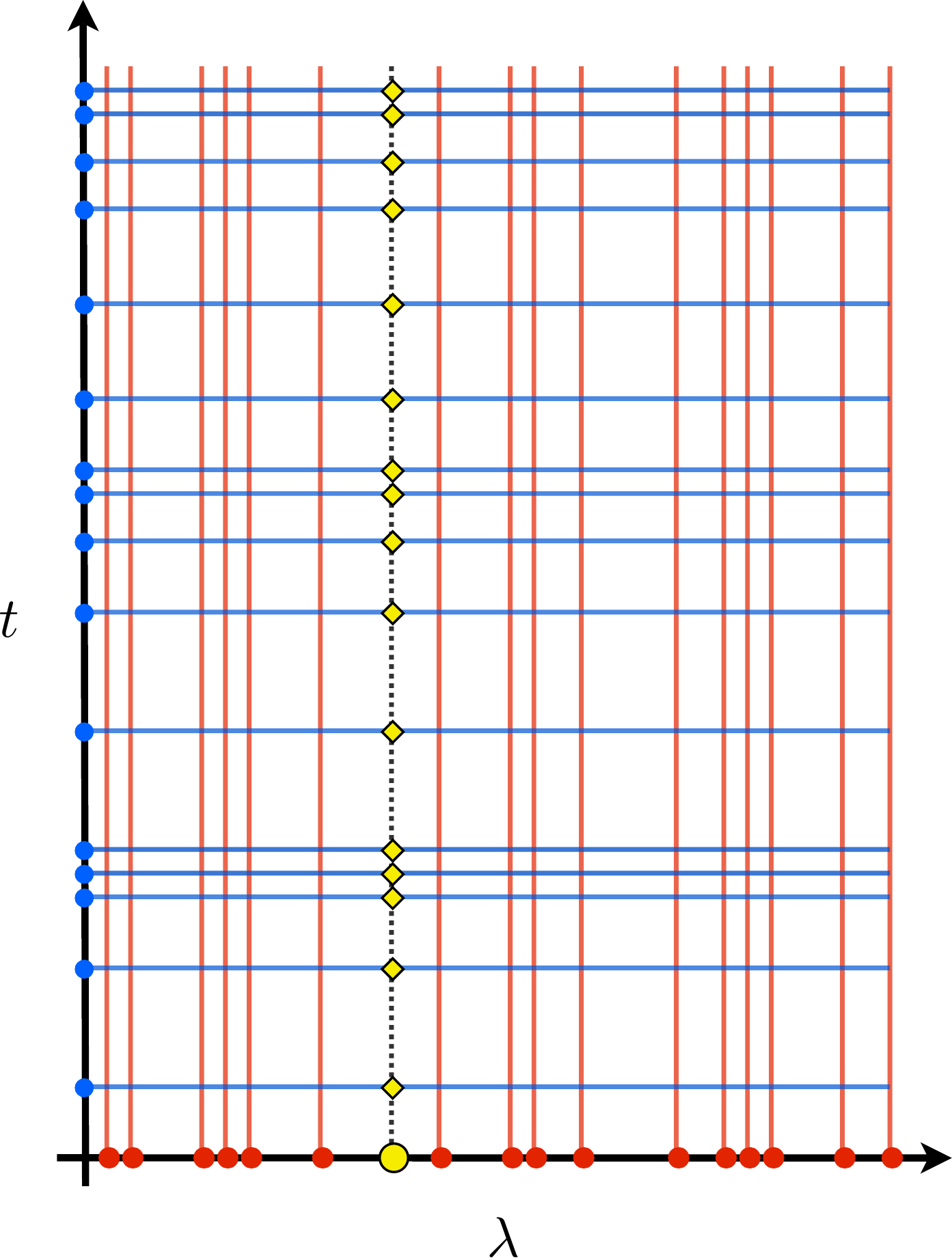}
\caption{A schematic of the method for building and evaluating the surrogate model. The red dots show the greedy selection of parameter points for building the reduced basis (Step 1, offline), the blue dots (Step 2, offline) show the associated empirical nodes in time from which a waveform can be reconstructed by interpolation with high accuracy, and the blue lines (Step 3, offline) indicate a 
fit for the waveform's parametric dependence at each empirical time. The yellow dot shows a generic parameter, which is predicted at the yellow diamonds and filled in between for arbitrary times using the empirical interpolant, represented as a dotted black line (Step 4, online).} 
\label{fig:SurrogateSchematic}
\end{figure}

The algorithm for building and evaluating a surrogate for a given fiducial family or model of gravitational waveforms is schematically depicted in Fig.~\ref{fig:SurrogateSchematic} and outlined below:
\begin{enumerate}
\item {\bf (Offline)} Described in Section~\ref{sec:Greedy}. Select the most relevant $m$ points in parameter space (shown as red dots in Fig.~\ref{fig:SurrogateSchematic}). The waveforms associated with these selections (shown as red lines) provide a nearly optimal reduced basis (RB) for this waveform family \cite{Field:2011mf}.
The resulting points and waveforms will be referred to as {\em greedy data}. 
\item  {\bf (Offline)} Described in Section~\ref{sec:EIM}. Identify $m$ time samples of the full time series, which we call {\it empirical nodes} or {\it times}, to build an interpolant that accurately reconstructs any fiducial waveform.
This step, called the Empirical Interpolation Method (EIM), only requires knowing the reduced basis. The number of empirical nodes $m$ (shown as blue dots on the vertical axis in Fig.~\ref{fig:SurrogateSchematic}) exactly equals the number of basis elements $m$.
\item  {\bf (Offline)} Described in Section~\ref{sec:LS}. At each empirical node perform a fit (e.g., least squares) in the parameter dimension for the amplitude and phase of the waveform using the greedy data from Step 1. 
The fits are indicated by blue lines in Fig.~\ref{fig:SurrogateSchematic}.
\item  {\bf (Online)} Described in Sec.~\ref{sec:complete_surrogate}. Evaluate the surrogate model constructed in Steps $1$-$3$ at any parameter value $\mlam_0$, shown as the yellow dot on the horizontal axis in Fig.~\ref{fig:SurrogateSchematic}. This is accomplished by computing the values of the amplitude and phase fits from Step 3 at each empirical node in time for $\mlam = \mlam_0$ (yellow diamonds). The full time series of the surrogate waveform is then generated using the empirical interpolant from Step 2 (dotted black vertical line).
\end{enumerate}

We quantify the accuracy of the offline steps through the convergence rates in (\ref{eq:greedyErr}) and (\ref{eq:errorEIM}). The accuracy of the fast online step for the complete surrogate is estimated through the errors in  (\ref{eq:l2_surrogate_error_d}) and (\ref{eq:pointwise}). 
If each offline step is carried out with sufficiently good accuracy
then the surrogate will satisfy 
\be
h_{\rm S}(t;\mlam) \approx h(t;\mlam) \label{eq:approx} 
\ee
for all $t$ and $\mlam$ in the given ranges and retain the physics of the original fiducial waveform family, whatever that might be.
As discussed in Sec.~\ref{sec:assess_surrogate}, the waveform predictions by our surrogate model are indeed expected to have a small error with respect to the fiducial one.

\section{Surrogate model building} \label{sec:SurrogateModels}

The following four subsections expand on the steps outlined above. Each of these steps is illustrated with an application to non-spinning EOB waveforms. For simplicity we consider the $(2,2)$ mode of waveforms with mass ratios in the range $q\in [1,2]$ and about $12,\!000 M$ in duration. In Sec.~\ref{sec:astro} we build surrogate models for astrophysical sources that include more cycles, cover larger mass ratio intervals, and contain higher spherical harmonics. Important technical details describing how these EOB waveforms were generated as well as our peak alignment scheme are discussed in Appendix~\ref{sec:EOB}.  Figure~\ref{fig:EOB1} shows the $q=1$ EOB waveform. Despite its complicated structure, we shall demonstrate that waveforms such as this one can be represented accurately by relatively little information.

A gravitational waveform $h(t; \mlam)$ is represented in terms of its two fundamental polarizations $h_+(t; \mlam)$ and $h_\times (t; \mlam)$ by $h(t; \mlam) = h_+(t; \mlam) + i h_\times (t; \mlam)$.
 A natural inner product is given by the complex scalar product
\be
\langle h(\cdot; \mlam_1),  h(\cdot; \mlam_2) \rangle = \int_{t_{\rm min}}^{t_{\rm max}} {\hskip-0.15in} dt \, h^*(t; \mlam_1) h (t; \mlam_2)  , \label{eq:dotproduct}
\ee
with an inherited norm given by
$\| h(\cdot; \mlam) \|^2 = \langle h(\cdot; \mlam), h(\cdot; \mlam)  \rangle$. 
Here, $h^*(t; \mlam)$ is the complex conjugate of $h(t; \mlam)$. 
Other inner products might be more natural for different applications \footnote{For data analysis applications other inner products may be natural. For matched filtering purposes, for example, the natural one would be the frequency domain overlap weighted by the inverse of the detector's power spectral density.}. Throughout this paper we shall assume the waveforms are normalized such that $\| h(\cdot; \mlam) \| = 1$.  

The overlap integral of two normalized waveforms, say, of a fiducial waveform and its surrogate model prediction, is given by ${\rm Re} \langle h(\cdot; \mlam), h_{\rm S}(\cdot; \mlam) \rangle$, 
\be
 {\rm Re} \langle h(\cdot; \mlam), h_{\rm S}(\cdot; \mlam) \rangle = 1 - \frac{1}{2} \| h(\cdot; \mlam) - h_{\rm S}(\cdot; \mlam) \|^2 \, . \label{eq:norm_overlap}
\ee
This equality is useful to translate the error in approximating a fiducial waveform by its surrogate model prediction into an overlap integral that is used in some gravitational wave applications (c.f., Eq.~\eqref{eq:mm}). 

\begin{figure} 
\includegraphics[width=0.98\linewidth]{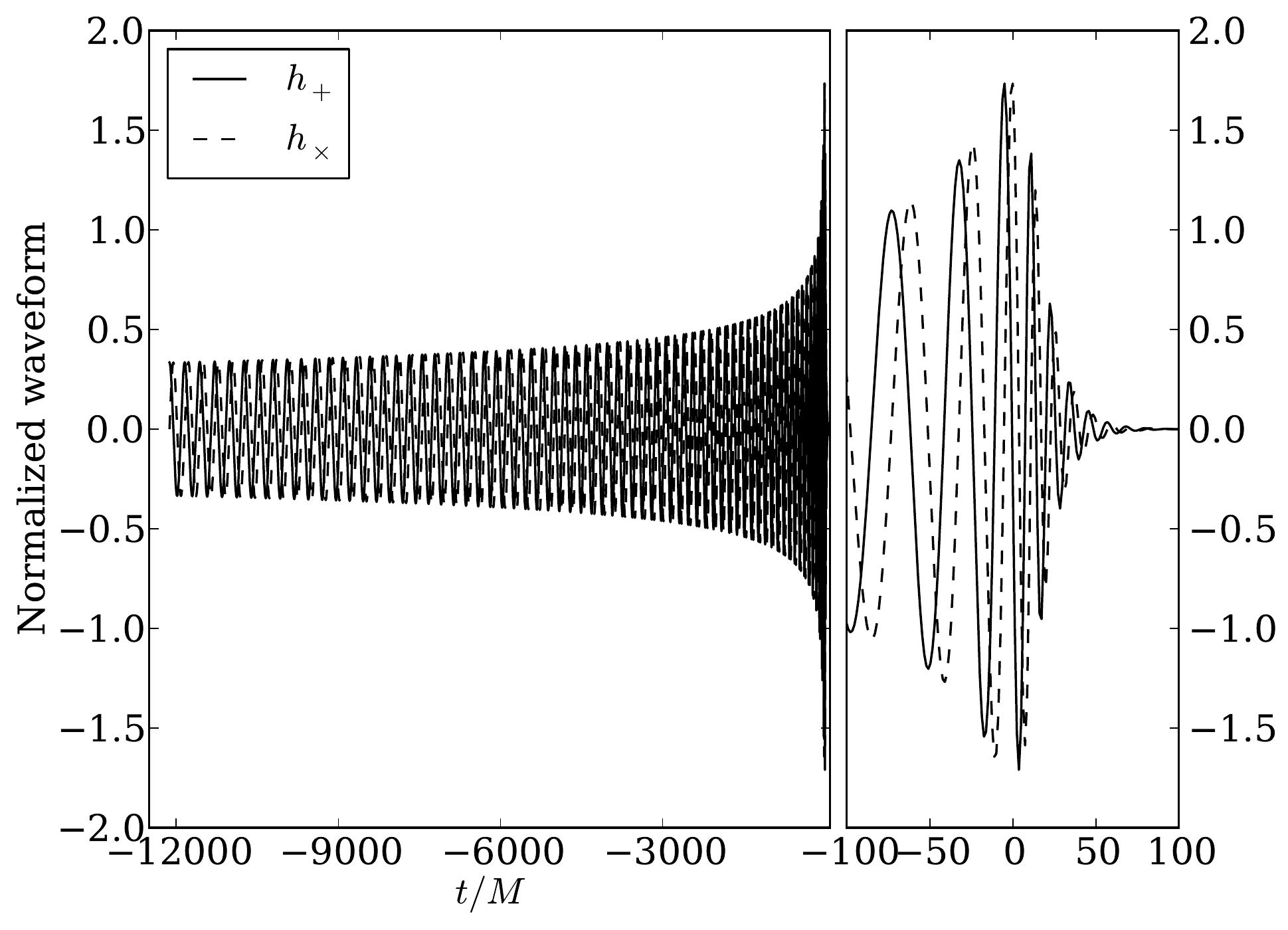}
\caption{Time series of a normalized $(2,2)$ mode of an EOB waveform for an equal mass, non-spinning black hole binary coalescence. This waveform, corresponding to about $70$ gravitational wave cycles, is representative of the structure encountered when building a surrogate model.
} 
\label{fig:EOB1}
\end{figure}

\subsection{Step 1: Greedy selection of parameter samples and reduced basis} \label{sec:Greedy}

We use a greedy algorithm (see Appendix~\ref{sec:RBapp} for more details) to select $m$ parameter points 
$\{ \mLam_i \}_{i=1}^m$
and corresponding waveforms $h_i(t) = {h(t;\mLam_i)}$. The greedy algorithm provides a nearly optimal solution to the  
Kolmogorov $n$-width approximation problem~\cite{Binev10convergencerates,DeVore2012}, namely, given a set of 
waveforms
\be
\{ h(t;\mlam) : \mlam \in {\cal T} \} \, ,
\ee
where ${\cal T}$ denotes a compact parameter domain,
find an $m$-dimensional function space that best approximates any $h(t;\mlam)$ from this set. 

More precisely, if the waveforms are known at a discrete set of $M$ training points
${\cal T}_M = \{\mlam_i \}_{i=1}^M$,
the greedy algorithm identifies a set of parameter values
 \be
\{ \mLam_1, \mLam_2, \ldots , \mLam_m \} \subset {\cal T}_M
\ee
and an associated set of waveforms 
\begin{align}
	\{ h_1(t), h_2(t), \ldots , h_m(t) \}
	\label{eq:basis100}
\end{align}
that constitutes the reduced basis. The basis is
hierarchical in the sense that if $\{h_i \}_{i=1}^{m'}$ is the basis for $m' < m$ then  
\be
\{ h_i \}_{i=1}^{m'} \subset \{h_i\} _{i=1}^m . 
\ee
One of the key features of the greedy algorithm is its ability to select a small number of waveforms to serve as an accurate basis. For practical purposes of conditioning 
it is useful to use an orthonormal basis
$\{ e_i \}_{i=1}^m$, which spans the same approximation space as (\ref{eq:basis100}).

With the RB in hand, every waveform in the training set is well approximated by an expansion of the form 
\be
h(t; \mlam ) \approx \sum_{i=1}^m c_i (\mlam ) e_i (t) \, , \label{eq:projapprox}
\ee
whereas waveforms from ${\cal T}$ (even if {\em not} in the training set) continue to be well approximated by 
the RB if the training set is dense enough~\cite{Field:2011mf,Caudill:2011kv,PhysRevD.86.084046,Cannon:2011xk}. Since the waveform space is numerically finite dimensional~\cite{Field:2011mf}, one can verify sufficiently dense training sets through convergence as $M$ gets larger or by checking how well the basis represents randomly selected waveforms (see Appendix \ref{sec:RBapp}). For an underlying model that requires prohibitively expensive numerical solves, one may use a simpler model to propose a training set building strategy. If there is sufficient similarity amongst the members of the original set then $m \ll M$. This is found to be the case for gravitational waveforms \cite{Field:2011mf,Caudill:2011kv,PhysRevD.86.084046}.

Let $\epsilon$ be a user specified tolerance whose role is to guarantee that the approximation error for waveforms in the training set, which we will call the {\it greedy error} $\sigma_m$, is bounded by $\epsilon$,
\begin{align} \label{eq:greedyErr}
\sigma_m \equiv \max_{\mlam } \min_{c_i \in \mathbb{C}} \bigg\| h(\cdot ;\mb{\lambda}) - \sum_{i=1}^m c_i(\mb{\lambda}) e_i(\cdot) \bigg\|^2 \leq \epsilon  \, .
\end{align}
Then,  the representation \eqref{eq:projapprox} is accurate to $\epsilon$. 
The minimization over the coefficients $\{ c_i \}$ in (\ref{eq:greedyErr}) is achieved by orthogonal projection ${\cal P}_m h(t;\mlam)$ of 
$h(t;\mb{\lambda})$ onto the span of the basis (see Appendix \ref{sec:RBapp} for details) so that
\begin{align} \label{eq:projcoeff}
c_i (\mlam) = \langle h(\cdot;\mlam) , e_i(\cdot) \rangle \, .
\end{align}
In Sec.~\ref{sec:EIM} we will find efficient approximations of the optimal projection representation in (\ref{eq:projcoeff}) that approximately retains its accuracy implied by (\ref{eq:greedyErr}).

The error in (\ref{eq:greedyErr}) is directly related to the overlap between a waveform and its representation \footnote{Strictly speaking, (\ref{eq:mm}) is an exact equality if ${\cal P}_m h$ is normalized. One can normalize it without loss of generality or, alternatively, take the point of view that for the high accuracies achieved by the reduced basis, (\ref{eq:mm}) should be seen as an approximate equality, at the level of machine precision in most cases.}
\be
\min_{\mlam} {\rm Re} \langle h(\cdot; \mlam ) , {\cal P} _m h (\cdot; \mlam ) \rangle = 1 - \frac{1}{2}\sigma_m  \, , \label{eq:mm}
\ee
which follows from \eqref{eq:norm_overlap}. 
 The quantity $\sigma_m$ quantifies the worst error of the best approximation by the basis. The greedy algorithm is nearly optimal in the sense that if the Kolmogorov $n$-width $d_m$ (defined as the smallest error (\ref{eq:greedyErr}) achieved by a best $m$-dimensional function space) decays exponentially then so does the greedy error \cite{Binev10convergencerates,DeVore2012}, 
\be
d_m \leq D \mathrm{e}^{-a   m^b }  \implies \sigma_m \leq \sqrt{2D} \mathrm{e}^{ -\tilde{a} m^b } \, , \label{eq:nwidth} 
\ee
where $D$, $a$, $b$ are positive constants and $\tilde{a}=2^{-1-2b}a$. 

Recent work~\cite{Field:2011mf,Caudill:2011kv,PhysRevD.86.084046,Cannon:2011xk} has shown that for fixed but arbitrary physical and parameter ranges, a small number of basis functions is indeed sufficient to accurately represent any waveform of the same physical model and with an exponentially decaying greedy error (\ref{eq:greedyErr}). Such observations are  expected for functions with smooth parameter dependence, as is the  case with gravitational waveforms. 
To better understand these approximation properties one can make an analogy to the more familiar case of spectral methods. There, exponential decay with the number of basis elements is expected whenever there is smoothness with the physical dimension(s) (e.g., space or time). 

\begin{figure}[ht]
\includegraphics[width=0.98\linewidth]{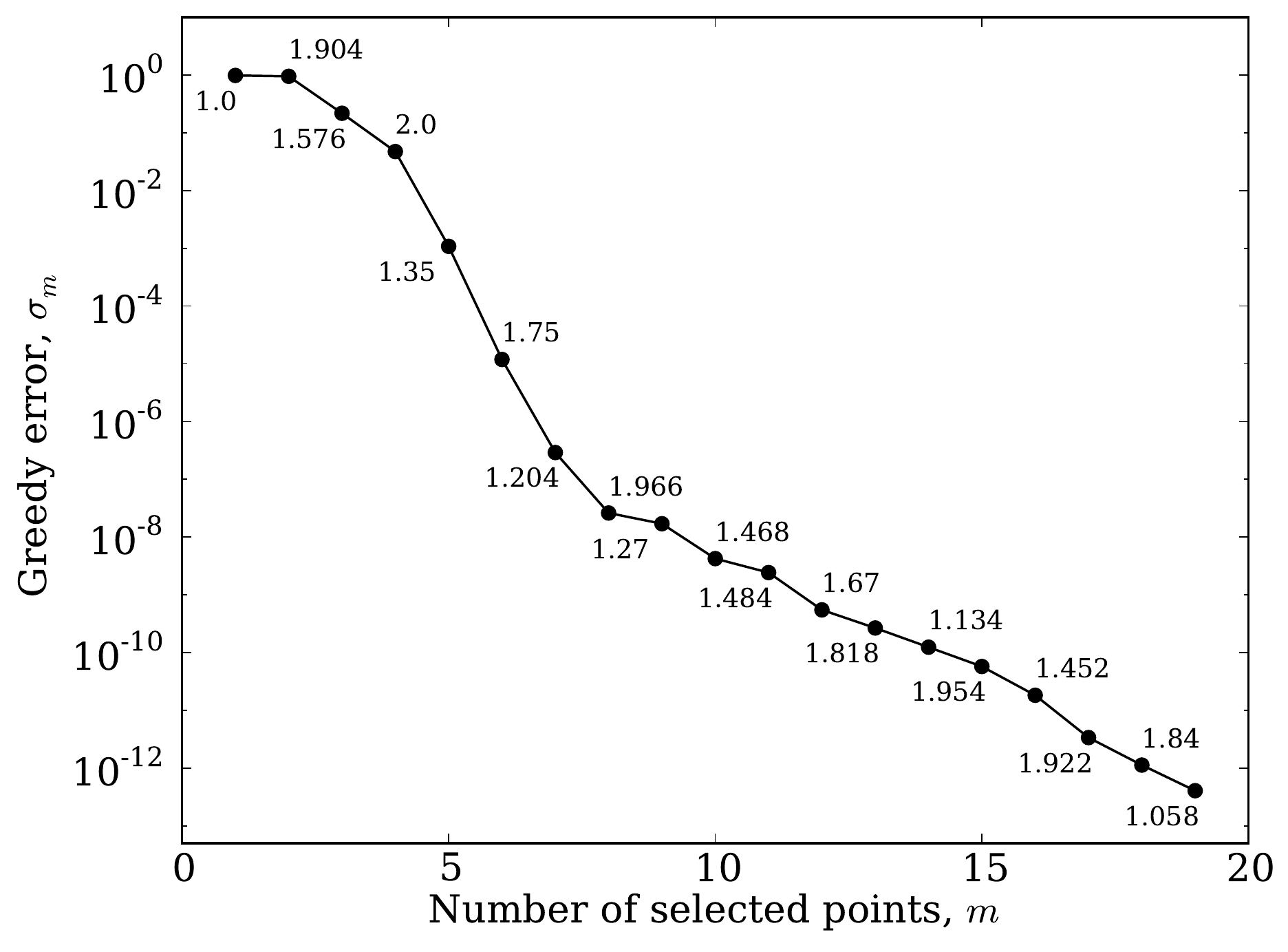}
\caption{Greedy error, as defined by (\ref{eq:greedyErr}), over $501$ EOB training set waveforms with mass ratios between $1$ and $2$. Labels at the dots indicate the selected mass ratios at each step in the greedy algorithm. 
} 
\label{fig:EOB_ExampleGreedy}
\end{figure}

\begin{figure}[ht]
\includegraphics[width=0.98\linewidth]{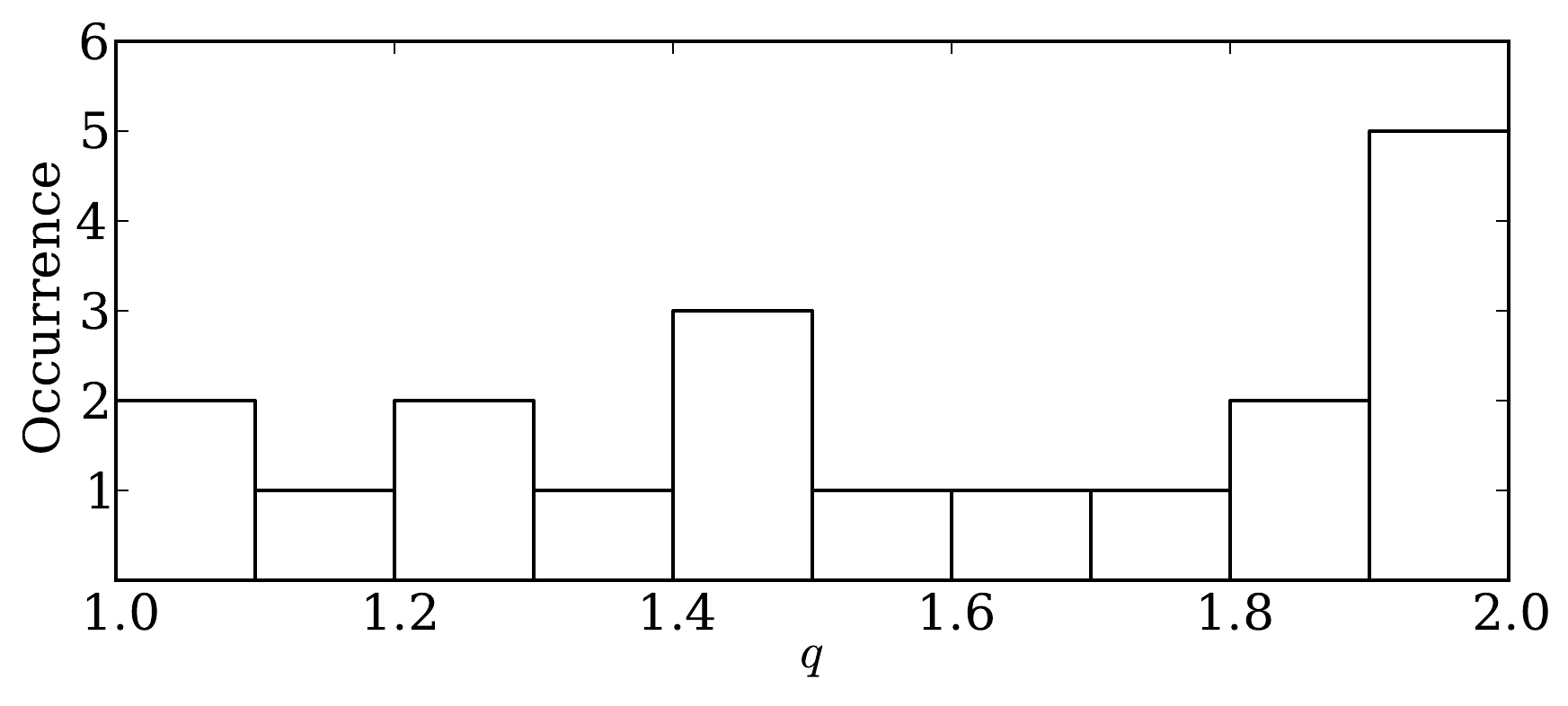}
\caption{Histogram of parameters selected by the greedy algorithm for the reduced basis of Fig.~\ref{fig:EOB_ExampleGreedy}.} 
\label{fig:EOB_ExampleParameters}
\end{figure}

Let us apply the greedy algorithm to build a reduced basis for our nominal EOB example introduced earlier.
Figure~\ref{fig:EOB_ExampleGreedy} shows the exponential decay of the greedy error (\ref{eq:greedyErr}) over $501$ waveforms in the training set, with only $19$ RB waveforms needed to represent the EOB model to machine precision for the mass ratios considered. Errors of about $10^{-3}$ are already achieved with as few as $5$ RB waveforms. 
Later on (c.f.,~Fig.~\ref{fig:EOB_ExampleSurrogateError}), we show that any waveform not present in the training set yields similarly small representation errors by the basis. This feature, due to a sufficiently well sampled training set, is essential for parameter estimation studies, which seek to explore the waveform continuum. The distribution of selected points is shown in Fig.~\ref{fig:EOB_ExampleParameters}. In Sec.~\ref{sec:LS} we show how the greedy data from these parameter selections can be used to {\em predict} waveforms for any $q$ in the range considered, including (and especially) values not in the original training set. 

\subsection{Step 2: Greedy selection of time samples and empirical interpolation} \label{sec:EIM}

Once a basis is built in Step 1 we can express any waveform evaluated at any time as a sum of $m$ reduced basis elements. In Step 2, which is shown to significantly reduce the surrogate's evaluation cost in Appendix~\ref{sec:donot}, we now show how to leverage this knowledge to yield a temporal prediction scheme by recasting the problem as one of interpolation in time. Given a reduced basis $\{e_i\}_{i=1}^m$ and $m$ evaluations of a fiducial waveform at certain times $\{T_i\}_{i=1}^m$, we wish to recover the full fiducial waveform $h(t; \mlam)$ with high accuracy for an arbitrary $\mlam$. A proper choice of these times $\{ T_i \} _{i=1}^m$ is crucial. Naively selected times, such as those randomly or equally spaced, do not guarantee that: i) the interpolation problem is well-conditioned or even has a solution, and ii) the interpolation error is minimized with a nearly optimal convergence rate.

A framework for finding a ``good'' set of times $\{T_i\}_{i=1}^m$ that achieve both criteria is provided by the Empirical Interpolation Method (EIM) \cite{Barrault2004667,Maday_2009}. These special times, which we call {\it empirical times} or {\it nodes}, are selected as a (sparse) subset of the waveform's given time series (or even the continuum). The empirical nodes are uniquely defined by the reduced basis waveforms and only these waveforms. Like the algorithm for building a reduced basis, the EIM is hierarchical and uses a greedy optimization strategy to select the most representative times. While the empirical times $T_i$ do not explicitly depend on parameters or their ranges, the parameter dependence is {\em implicit}, nevertheless, through the basis. For example, a reduced basis for spinning or precessing waveforms will exhibit different features, and the distribution of $T_i$ will reflect this. For the moment we shall assume that the empirical nodes are known; the precise algorithm for finding them is given in Appendix \ref{sec:EIMapp}.

The empirical interpolant, which interpolates the waveform $h(t; \mlam)$ in time for a given parameter $\mlam$, is denoted by ${\cal I}_m [h](t;\mlam)$ and takes the form
\be
{\cal I}_m [h](t;\mlam) = \sum_{i=1}^m  C_i (\mlam) e_i(t)\, .  \label{eq:interpdef}
\ee
The coefficients $\{ C_i \}_{i=1}^m$ are defined by requiring the interpolant to equal the value of the waveform at the empirical nodes,
\begin{align} \label{eq:intproblem} 
\sum_{i=1}^m  C_i (\mlam) e_i(T_j) = h(T_j; \mlam) , \qquad  j=1,\dots, m \, , 
\end{align}
which is equivalent to solving an $m$-by-$m$ system 
\be
\sum_{i=1}^mV_{ji}C_i(\mlam) = h (T_j; \mlam )\,, \quad j=1, \ldots, m
\ee
for the coefficients $\{ C_i \}_{i=1}^m$ where the interpolation matrix
\begin{equation} \label{eq:InterpMatrix}
  V \equiv \left(  \begin{array}{cccc}   
              e_1(T_1)  &  e_2(T_1)            & \cdots & e_{m}(T_1)      \\
              e_1(T_2)  &  e_2(T_2)            & \cdots & e_{m}(T_2)       \\
              e_1(T_3)  &  e_2(T_3)          & \cdots & e_{m}(T_3)   \\              
              \vdots    & \vdots             & \ddots & \vdots                       \\
              e_1(T_{m})  & e_2(T_{m})    & \cdots & e_{m}(T_{m})  \\               
             \end{array}
   \right) 
\end{equation} 
is independent of the parameters $\mlam$. 

The choice of empirical nodes given by the EIM algorithm
together with the linear independence of the reduced basis ensure that $V$ in (\ref{eq:InterpMatrix}) is as well-conditioned as possible and invertible~\footnote{Linear independence of the basis functions does not imply linear independence of vectors which come from evaluations of those functions.} so that 
\be
C_i = \sum_{j=1}^m \left( V^{-1}\right)_{ij}h (T_j; \mlam)   \label{eq:CEIM}
\ee
is the unique solution to (\ref{eq:intproblem}).  
It then follows upon substituting (\ref{eq:CEIM}) into (\ref{eq:interpdef}) that the empirical interpolant is
\be \label{eq:EIM_with_B}
{\cal I}_m [h](t;\mlam) = \sum_{j=1}^m B_j (t) h(T_j; \mlam )
\ee 
where 
\be
B_j(t) \equiv  \sum_{i=1}^m e_i (t )  \left( V^{-1} \right)_{ij}    \label{eq:BEIM}
\ee
and is independent of $\mlam$.
Note that (\ref{eq:EIM_with_B}) is  a linear combination of the fiducial waveform itself evaluated at the empirical times.
The coefficients $\{ B_i \}_{i=1}^m$ are built directly from the reduced basis and provide a clean offline/online separation. Because of this the $\{B_i\}_{i=1}^m$ can be pre-computed offline once the reduced basis is generated while the (fast) interpolation is computed during the online stage from (\ref{eq:EIM_with_B}) when the parameter $\mlam$ is specified by the user. Evaluations of the fiducial waveform are still needed at the arbitrarily chosen parameter $\mlam$ in order to construct the interpolant in (\ref{eq:EIM_with_B}). In the next subsection we explain how to estimate the fiducial waveform at any $\mlam$, thus approximating $\{h(T_i; \mlam)\}_{i=1}^m$ and completing the construction of the surrogate model.

The empirical interpolant satisfies \cite{antil2012two}
\begin{align}
\max_{\mlam } \| h(\cdot ;\mb{\lambda}) - {\cal I}_m [h](\cdot ;\mb{\lambda}) \|^2 \leq \Lambda_m \sigma_m \, , \label{eq:errorEIM}
\end{align}
where $\sigma_m$ is the greedy error defined in \eqref{eq:greedyErr} and $\Lambda_m$ is a computable Lebesgue-like quantity that changes slowly with $m$ (see Appendix \ref{sec:EIMapp}). For problems with smooth dependence with respect to parameter variations we can expect an exponential decay of $\sigma_m$ with $m$ and of the empirical interpolant's error.

\begin{figure}[ht]
\includegraphics[width=\columnwidth]{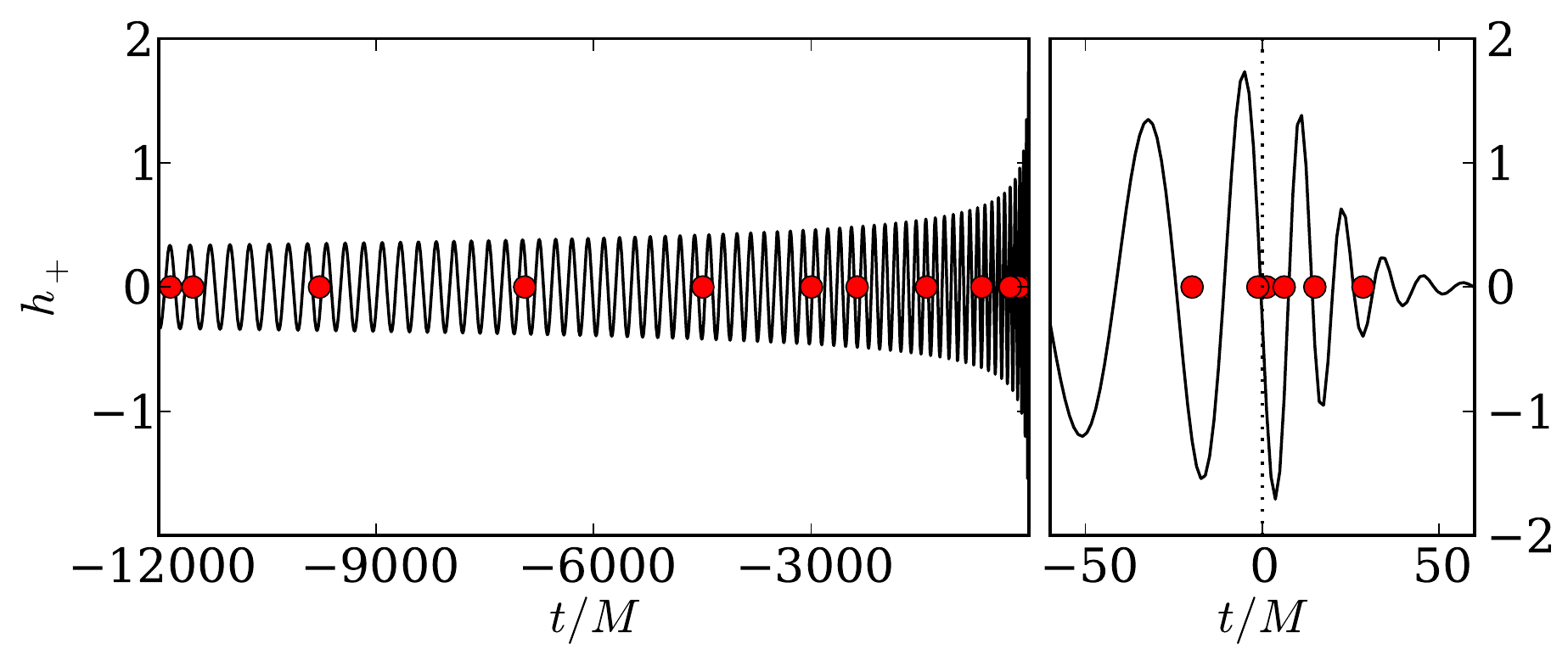}
\caption{Location of the empirical nodes for the fiducial family of EOB waveforms with mass ratio $q\in[1,2]$. Knowing the waveform in this parameter range at these specific times is sufficient to reconstruct the former with very high accuracy at any other time using the empirical interpolant in \eqref{eq:EIM_with_B}.} 
\label{fig:EOB_ExampleDEIM}
\end{figure}

\begin{figure}[ht]
\includegraphics[width=0.98\linewidth]{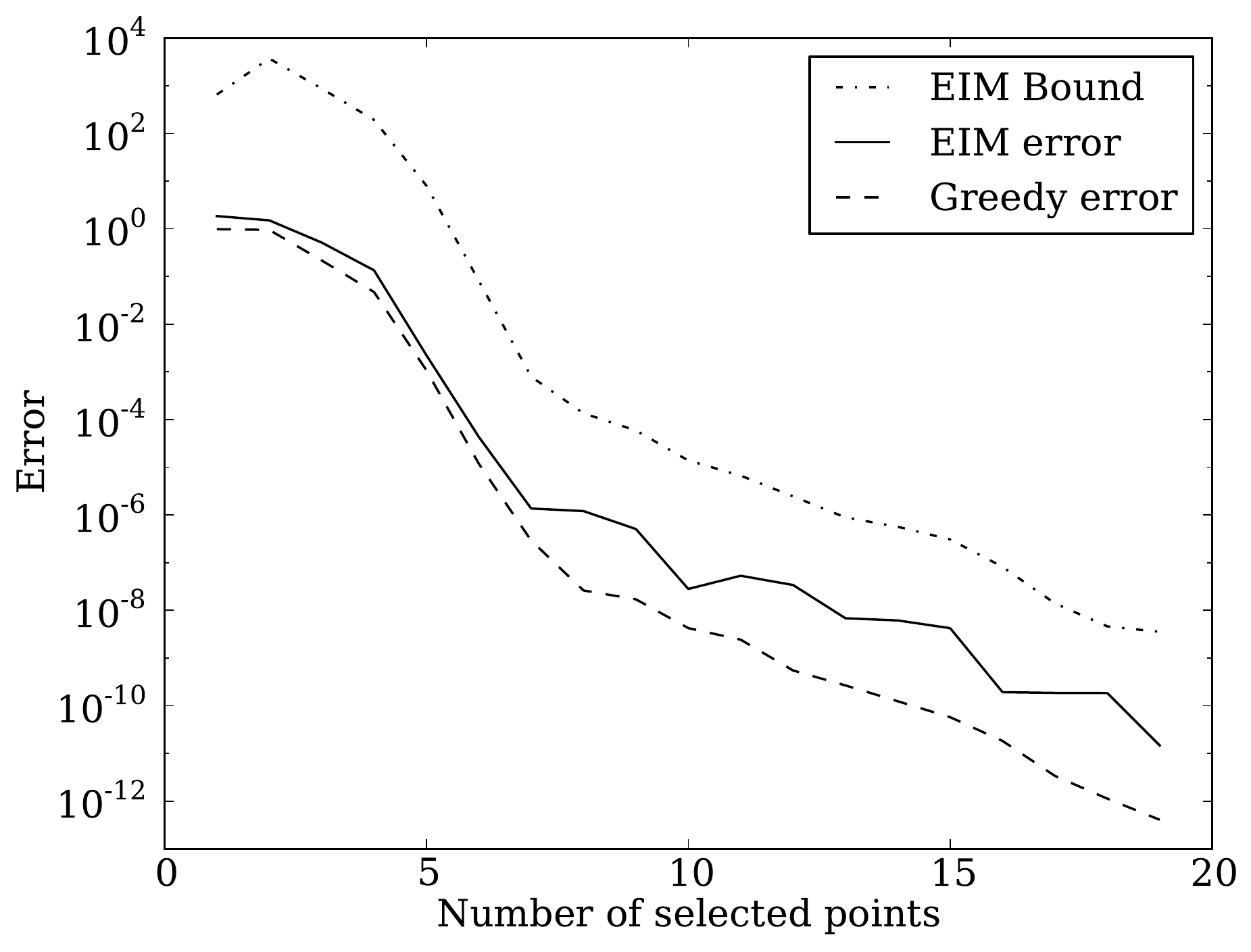}
\caption{A comparison of errors for the example family of EOB waveforms. The dashed line shows the greedy error $\sigma_m$ in (\ref{eq:greedyErr}). The solid line shows the maximum empirical interpolant error (\ref{eq:EIMerrMC}) taken over $1,\!000$ randomly selected waveforms (i.e., not taken from the training set) for $q \in [1,2]$. 
The dash-dotted line shows the error bound provided by the right side of (\ref{eq:errorEIM}) and is based solely on the greedy error and $\Lambda_m$. All three errors display similar decay rates.
} 
\label{fig:EOB_ExampleDEIMErrs}
\end{figure}

Before describing how to estimate the values $\{h(T_i; \mlam)\}_{i=1}^m$ for arbitrary $\mlam$ let us assume these values are known exactly and apply the EIM to build an empirical interpolant for our fiducial EOB example introduced earlier. Figure~\ref{fig:EOB_ExampleDEIM} shows all $19$ empirical nodes set against a $q=1$ waveform to compare with the structure of a typical waveform. Evaluating any $q\in[1,2]$ EOB waveform at these $19$ nodes and computing (\ref{eq:EIM_with_B}) one can reconstruct the full time-series of the waveform with high accuracy. 
This is explicitly demonstrated in Fig.~\ref{fig:EOB_ExampleDEIMErrs} where the solid black line denotes the largest empirical interpolation error 
\be
\| h(\cdot ;q) - {\cal I}_m [h](\cdot ;q) \|^2 \label{eq:EIMerrMC}
\ee
as a function of the number of reduced basis elements/empirical nodes
for $1,\!000$ randomly selected EOB waveforms drawn from $q\in[1,2]$. Notice that this error is remarkably close to the greedy error (dashed line) in (\ref{eq:greedyErr}) when using (\ref{eq:projcoeff}) for the coefficients. 
The bound in \eqref{eq:errorEIM} (dashed-dotted line) guarantees an error better than $10^{-8}$, which is sufficient for many GW applications. 

\subsection{Step 3: Fitting at empirical nodes} \label{sec:LS}

The next step is to predict waveforms at the empirical nodes $\{T_i \}_{i=1}^m$ for {\em arbitrary} parameter values $\mlam$ based only on the knowledge of the fiducial waveforms at the greedy points $\{ \mLam_i \}_{i=1}^m$.  
To accomplish this, we fit $h(T_i; \mlam)$ with respect to $\mlam$ at each $T_i$ using only the following $m$ values of the reduced basis waveforms:
\be
\{h(T_i; \mLam_j) \} _{j =1}^m \, .
\ee
The accuracy of the fit using only this data relies, at least partially, on the fact that the reduced basis waveforms are chosen to be the most dissimilar from one another. Of equal importance is our choice of fitting function which, in principle, is arbitrary. We will focus on the choices most effective for our nominal EOB example while others could be more appropriate for different waveform families.

The behavior of most astrophysically relevant gravitational waveforms is highly oscillatory in time but the phase and amplitude themselves have a relatively simple structure.  It is thus easier to perform high-accuracy fits of the phase and amplitude than of the complex waveform itself.
The amplitude $A$ and phase $\phi$ are defined through  
\be \label{eq:AmpPhase}
h(t;\mlam) = A(t;\mlam)e^{-i \phi(t;\mlam)} \, . 
\ee
This third step then consists of finding 
$2m$ functions, $\{A_i(\mlam)\}_{i=1}^m$  
and $\{\phi_i(\mlam)\}_{i=1}^m$, 
approximating the amplitude and phase of the waveform. Once these fitting functions have been found the approximation at each $T_i$ is
\begin{align}  \label{eq:LSh}
h(T_i; \mlam) \approx  A_i(\mlam) e^{-i \phi_i(\bm{\lambda})} \, . 
\end{align} 

Depending on the application some fitting functions might be more useful than others. Therefore, this third step in constructing a surrogate model is flexible in the way that the fitting is implemented and thus in how the surrogate is ultimately generated. This is quite a useful feature of the method that may be especially beneficial for building surrogate waveforms for highly precessing black hole binaries. Splines, rational polynomial, or weighted non-oscillatory fitting approaches could help limit the impact of numerical noise, for example.

\begin{figure}[ht]
\includegraphics[width=0.98\linewidth]{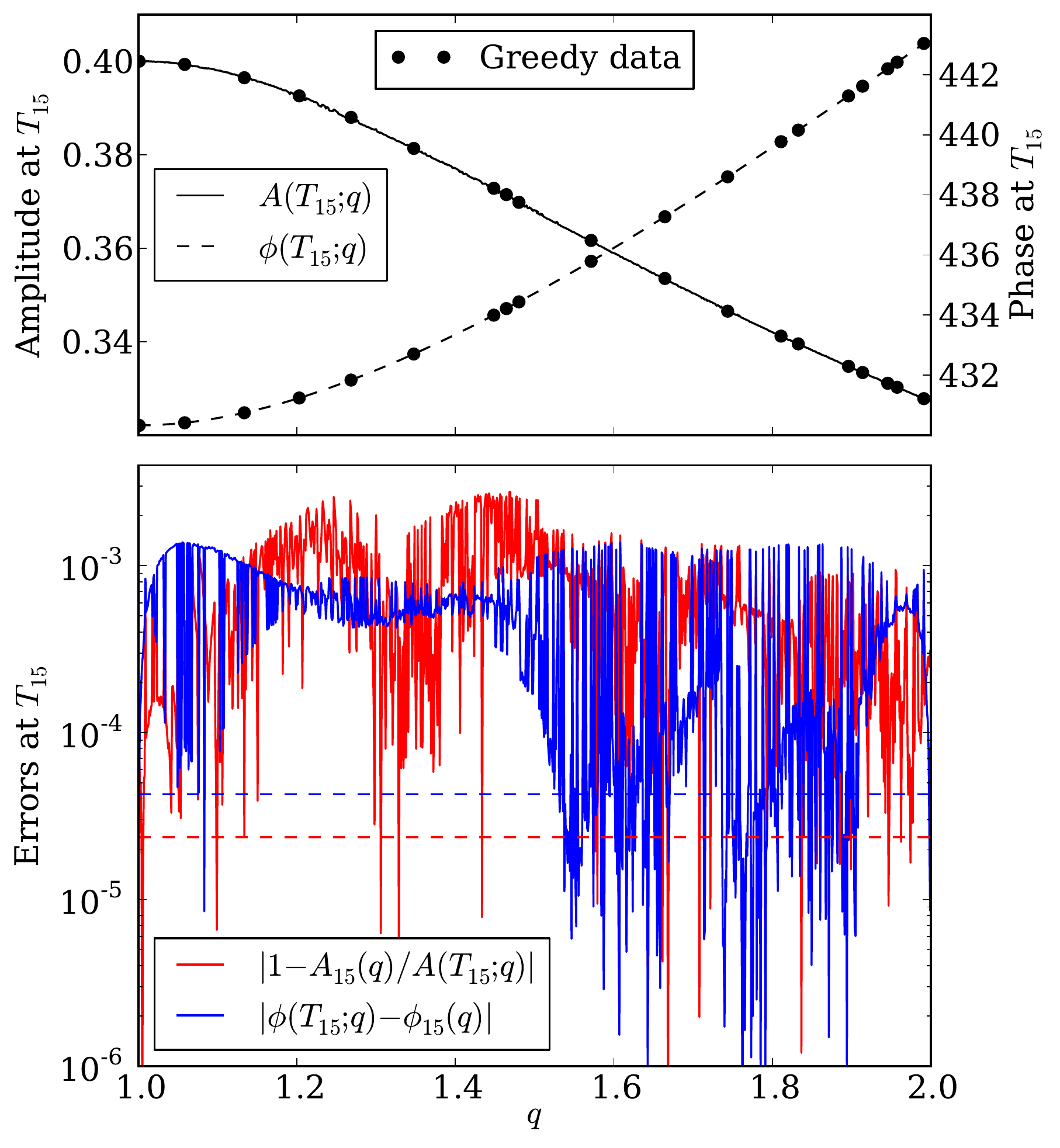} 
\caption{ 
{\bf Top}: Amplitude (solid) and phase (dashed) of the fiducial EOB training space waveform at the fifteenth selected empirical time as a function of $q$ along with the greedy data (circles). This empirical time is $T_{15} = 28.5M$ after merger and corresponds to the largest pointwise relative error for the least squares fit to the amplitude as quantified by (\ref{eq:pointwiseAmp}).
{\bf Bottom}: The pointwise least squares errors for the amplitude (red) and phase (blue) at $T_{15}$ evaluated for $1,\!000$ randomly selected waveforms. The dashed lines correspond to the maximum pointwise error for the second empirical node $T_2 = -2,\!367 M$, which has the smallest maximum error of all the nodes.
} 
\label{fig:EOB_ExampleAmpPhase}
\end{figure}

We now return to our nominal EOB example and perform a least squares fit for both the amplitude and phase as a function of mass ratio at each empirical time using polynomials,
\begin{align}
	A_i (q) = \sum_{n=0}^{\alpha_i} a_{i,n} q^n ~, ~~ \phi_i (q) = \sum_{n=0} ^{\beta_i} b_{i,n} q^n
\end{align}
where $\alpha_i, \beta_i < m$ are the degrees of the polynomials at the empirical time $T_i$ for $i=1,2, \ldots, m$. Further details regarding how to select an optimal degree are provided in Appendix~\ref{sec:polyLS}.

The top plot in Fig.~\ref{fig:EOB_ExampleAmpPhase} shows the amplitude and phase, along with the greedy data points, at the fifteenth empirical time node, $T_{15}$, which is about $28.5M$ after merger. This node corresponds to the largest pointwise error for the relative amplitude
\be \label{eq:pointwiseAmp}
\left | \frac{A(T_i; q) - A_i(q)}{A(T_i; q)} \right | 
\ee
for waveforms in the training set of our EOB test problem. $T_{15}$ also happens to correspond to the second largest difference for the phase, 
\be \label{eq:pointwisePhase}
| \phi(T_i; q) - \phi_i(q) | \, . \quad 
\ee
The bottom plot in Fig.~\ref{fig:EOB_ExampleAmpPhase} shows the pointwise errors (solid lines) of (\ref{eq:pointwiseAmp}) and (\ref{eq:pointwisePhase}) as a function of mass ratio for $1,\!000$ randomly selected waveforms. These errors are uniformly below $3 \times 10^{-3}$. The horizontal dashed lines show the maximum errors for the empirical node for which (\ref{eq:pointwiseAmp}) and (\ref{eq:pointwisePhase}) are smallest, which occurs for the second empirical time $T_2 = -2,\!367 M$. These errors are of order $10^{-5}$. As we will discuss later on (see Fig.~\ref{fig:EOB_ExampleSurrogateError}) all of this information translates into a mismatch of the surrogate model with respect to the underlying EOB family of $< 10^{-7}$. 

\begin{figure}[ht]
\includegraphics[width=0.98\linewidth]{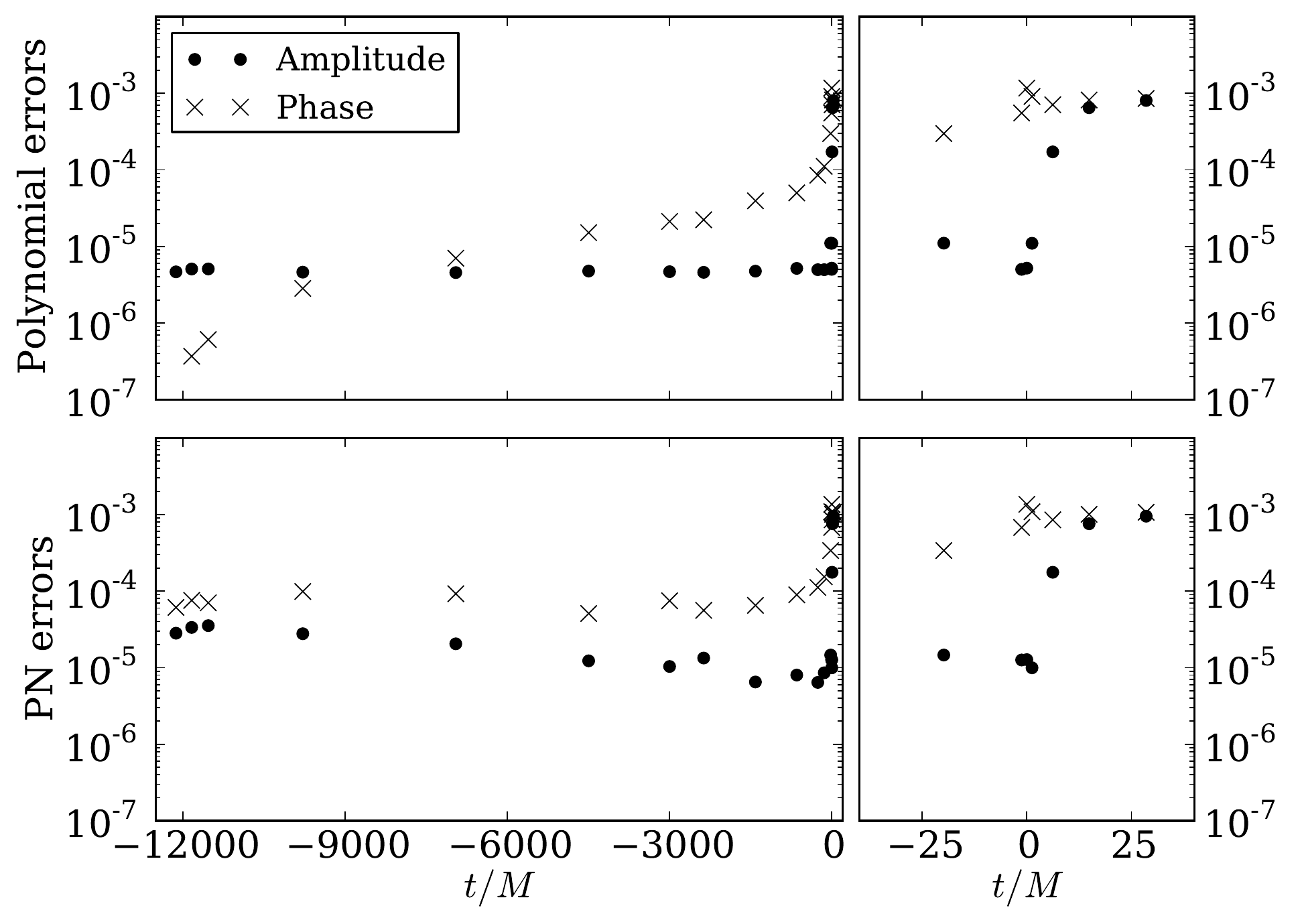}
\caption{The relative amplitude differences and phase differences of the least squares fits, as defined by (\ref{eq:pointwiseAmp}) and (\ref{eq:pointwisePhase}), maximized over the greedy mass ratios at each empirical time for our EOB example. The top panel shows these errors when using a polynomial least squares fit and the bottom panel when using a fitting function inspired by the post-Newtonian amplitude and phase.
Both types of fits exhibit very low errors at all of the empirical times.
}
\label{fig:EOB_ExampleGoodnessFit}
\end{figure}

The quality of a fit  at each empirical node, using the greedy data, depends on the smoothness of those waveforms with respect to parameter variation. This is discussed in Appendix~\ref{sec:EOB}. Here, it suffices to mention that the fitting errors depend sensitively on accurately aligning the waveforms at their peaks, which affects the fits most noticeably through merger and ringdown. This can be seen in the top panel of Fig.~\ref{fig:EOB_ExampleGoodnessFit}. 

Figure \ref{fig:EOB_ExampleGoodnessFit} shows the maximum of the pointwise differences from (\ref{eq:pointwiseAmp}) and (\ref{eq:pointwisePhase}) for the relative amplitude (circles) and phases (crosses), respectively, evaluated at each empirical time. We see that the amplitudes are accurate to better than $10^{-5}$ for the entire inspiral phase until the merger regime where the error increases to about $10^{-3}$ after which it plateaus throughout the ringdown stage. The phase errors increase modestly during the inspiral and likewise plateau through ringdown with errors at the level of $10^{-3}$.

Instead of using polynomials for the fitting functions we next consider functions inspired by the expressions for the amplitude and phase through leading order and next-to-leading order, respectively, in the post-Newtonian expansion
\begin{align}
	A_i (q) = {} & a_{i,0} \frac{ (q-1)^{a_{i,1} } }{ q^{a_{i,2}} } + a_{i,3} \\
	\phi_i (q) = {} & a_{i,0} \frac{ (q-1)^{a_{i,1} } }{ q^{a_{i,2}} } \big( 1 + a_{i,4} (q+ a_{i,5})^{a_{i,6} } \big) + a_{i,3} \,.
\end{align}
The bottom panel of Fig.~\ref{fig:EOB_ExampleGoodnessFit} shows the maximum of the pointwise differences from (\ref{eq:pointwiseAmp}) and (\ref{eq:pointwisePhase}) using these post-Newtonian-inspired fitting functions. These fitting functions have a least squares fitting error comparable to the polynomial errors shown in the top panel. In both cases, the fit quality decreases rapidly at the merger but still exhibit very low errors at all of the empirical times. We thus see in this example that the third offline step for building the surrogate is flexible in the choice of fitting functions. This insight could be useful for other fiducial models such as waveforms with precession. 

\subsection{Step 4: Completing the surrogate model} \label{sec:complete_surrogate}

Finally, our complete surrogate model $h_{\rm S}( t; \mlam)$ for the fiducial waveform family is given by substituting the fitting approximation (\ref{eq:LSh}) into the empirical interpolant ~(\ref{eq:EIM_with_B}), which yields
\be
	h_{\rm S}(t; \mlam) \equiv \sum_{i=1}^m B_i (t) A_i (\mlam) e^{-i \phi_i (\bm{\lambda})} \, . 
\label{eq:surrogate100}
\ee
This is the culmination of the offline steps. Only the $m$ reduced basis waveforms evaluated at the $m$ empirical times are needed to build the surrogate model and to predict an approximation for a fiducial waveform at any time and parameter value. In addition, the $\{B_i(t)\}_{i=1}^m$ are computed once and for all offline; only the fitting functions for the amplitude and phase need to be evaluated during the online stage once $\mlam$ is specified.

\section{Assessing the surrogate model} \label{sec:assess_surrogate}

One of the errors of interest for the complete surrogate model is a discrete version of the normed difference between a fiducial waveform and its surrogate, 
which is, for $L$ equally spaced time samples,
\be
 \Delta t \sum_{i=1}^L\left| h(t_i; \mlam) - h_{\rm S}(t_i; \mlam)\right|^2 \, , \label{eq:l2_surrogate_error_d}
\ee
where $\Delta t = (t_\mathrm{max} - t_\mathrm{min})/(L-1)$. We will sometimes refer to this as the {\it surrogate error}.
Recall, from (\ref{eq:norm_overlap}) and (\ref{eq:mm}) that the square of the normed difference between two waveforms is directly related to their overlap.
Other errors of interest are the pointwise ones for the phase and amplitude, 
\be \label{eq:pointwise}
\left | \frac{A(t; \mlam) - A_{\rm S}(t; \mlam)}{A(t; \mlam)} \right | \, , \quad | \phi(t; \mlam) - \phi_{\rm S}(t; \mlam) |  \, . 
\ee

\begin{figure}[ht]
\includegraphics[width=\columnwidth]{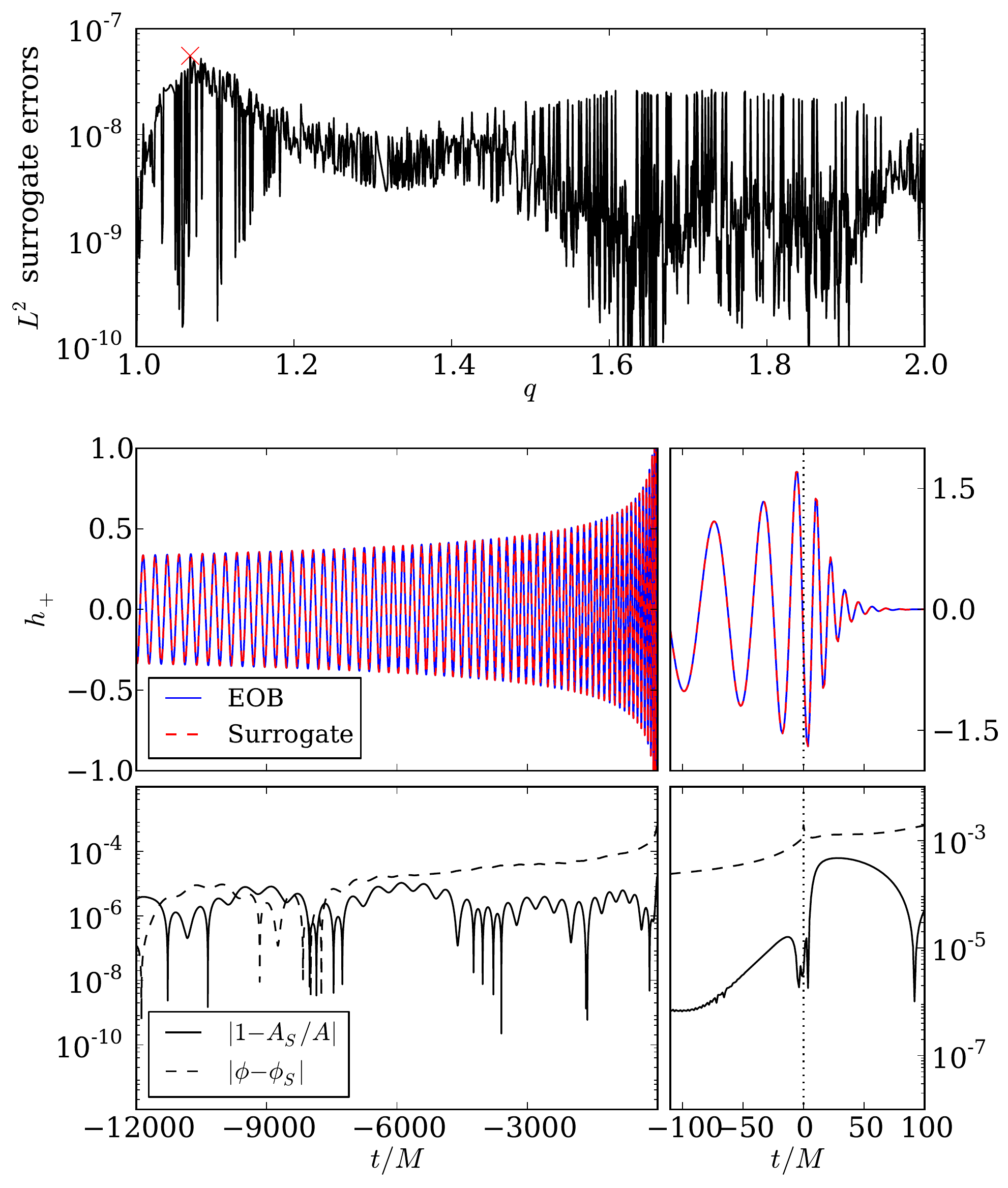}
\caption{
{\bf Top}: Surrogate model error defined by (\ref{eq:l2_surrogate_error_d}), which is related to the overlap error through (\ref{eq:norm_overlap}), for $1,\!000$ randomly selected mass ratios. The mass ratio yielding the largest surrogate model error is $q = 1.068$.
{\bf Middle}: The fiducial EOB waveform and its surrogate prediction for $q=1.068$. There is visual agreement throughout the entire duration of $\approx 12,\!000M$.
{\bf Bottom}: The fractional errors (\ref{eq:pointwise}) in the amplitude and the phase difference between the fiducial EOB waveform and its surrogate model prediction for $q=1.068$. The differences are smaller than the errors intrinsic to the EOB model itself as well as those of state-of-the-art numerical relativity simulations.
}
\label{fig:EOB_ExampleSurrogateError}
\end{figure}

Figure~\ref{fig:EOB_ExampleSurrogateError} shows a variety of comparisons between the surrogate and fiducial model for our EOB test case, using $L=16,\!384$ time samples \footnote{See Appendix~\ref{sec:EOB} for details about EOB waveform generation.}. The top plot shows that the surrogate error (\ref{eq:l2_surrogate_error_d}) is uniformly below $10^{-7}$, where the mass ratio $q =1.068$ corresponds to the largest error. The middle panel of Fig.~\ref{fig:EOB_ExampleSurrogateError} shows the fiducial EOB and surrogate waveforms for $q=1.068$. Both waveforms are visually indistinguishable and, from the bottom panel of the same figure, we see that both amplitude and phase pointwise errors (\ref{eq:pointwise}) are indeed very small. The largest errors are $\lesssim 10^{-3}$ and are smaller than: i) the differences, for the same quantities, between the EOB model and the NR simulations used to calibrate the former \cite{Pan:2011gk}, and ii) the numerical error of those NR simulations (see, e.g., \cite{Buchman:2012dw}) and of more recent state-of-the-art simulations \cite{Hinder:2013oqa}, as quantified through self-convergence tests. As discussed in Sec.~\ref{sec:LS} and App.~\ref{sec:EOB}, these maximum errors for the surrogate take place shortly after merger and are directly related to the accuracy with which one can determine the peak amplitude of the fiducial waveforms used to build the surrogate. 

In Appendix~\ref{sec:surrogateErrs} we derive the following error bound for the discrete norm (\ref{eq:l2_surrogate_error_d}),  
\begin{align} 
	{\hskip-0.5in}  \Delta t & \sum_{i=1}^L\left| h(t_i; \mlam) - h_{\rm S}(t_i; \mlam)\right|^2  \nonumber \\  
 & \leq \Lambda_m \sigma_m +  \Lambda_m \Delta t \sum_{i=1}^m \big( h(T_i, \mlam) - h_{\rm S}(T_i, \mlam) \big)^2  . 
 \label{eq:sErrorb}    
\end{align} 
This bound identifies contributions from two sources. The first term in (\ref{eq:sErrorb}) describes how well the empirical interpolant (i.e., the basis and empirical nodes) represents $h(t;\mlam)$. The expected exponential decay of the greedy error $\sigma_m$ with $m$ along with a slowly growing Lebesgue constant $\Lambda _m$ results in this term being very small. The term $\Lambda_m \sigma_m$ corresponds exactly to the curve labeled ``EIM Bound'' in Fig.~\ref{fig:EOB_ExampleDEIMErrs}. The second term in (\ref{eq:sErrorb}) is related to the quality of the fit. Incidentally, the fitting step has the dominant source of error in the surrogate model compared to the first two steps of generating the reduced basis and build the empirical interpolant (see also the discussion in Section \ref{sec:LS}).

\section{Cost and speedup for surrogate model predictions}
\label{sec:cost_speedup}

Next we discuss the cost (in terms of operation counts) to evaluate a surrogate model. We also present the large speedups that can be achieved when evaluating a surrogate model for our nominal EOB example compared to generating a fiducial waveform using the EOB solver as implemented in the LAL software package, which we refer to as the EOB-LAL code.
 
The complete surrogate model is given in (\ref{eq:surrogate100}) where the $m$ coefficients $B_i(t)$ in \eqref{eq:BEIM} and the $2m$ fitting functions $\{ A_i (\mlam) \}_{i=1}^m$ and $\{ \phi_i (\mlam) \}_{i=1}^m$ are assembled offline as described in Sections \ref{sec:Greedy}, \ref{sec:EIM}, and \ref{sec:LS}. 
In order to evaluate the surrogate model for some parameter $\mlam_0$ we only need to evaluate each of those  $2m$ fitting functions at $\mlam_0$, recover the $m$ complex values $\{ A_i(\mlam_0) e^{-i \phi_i(\mlam_0)} \}_{i=1}^m$, and finally perform the summation in \eqref{eq:surrogate100}.  
Each $B_i(t)$ is a complex-valued time series with $L$ samples.
Therefore, the overall operation count to evaluate the surrogate model at each $\mlam_0$ is $(2m-1)L$ 
plus the cost to evaluate the fitting functions. 

\begin{figure}[ht]
\includegraphics[width=0.98\linewidth]{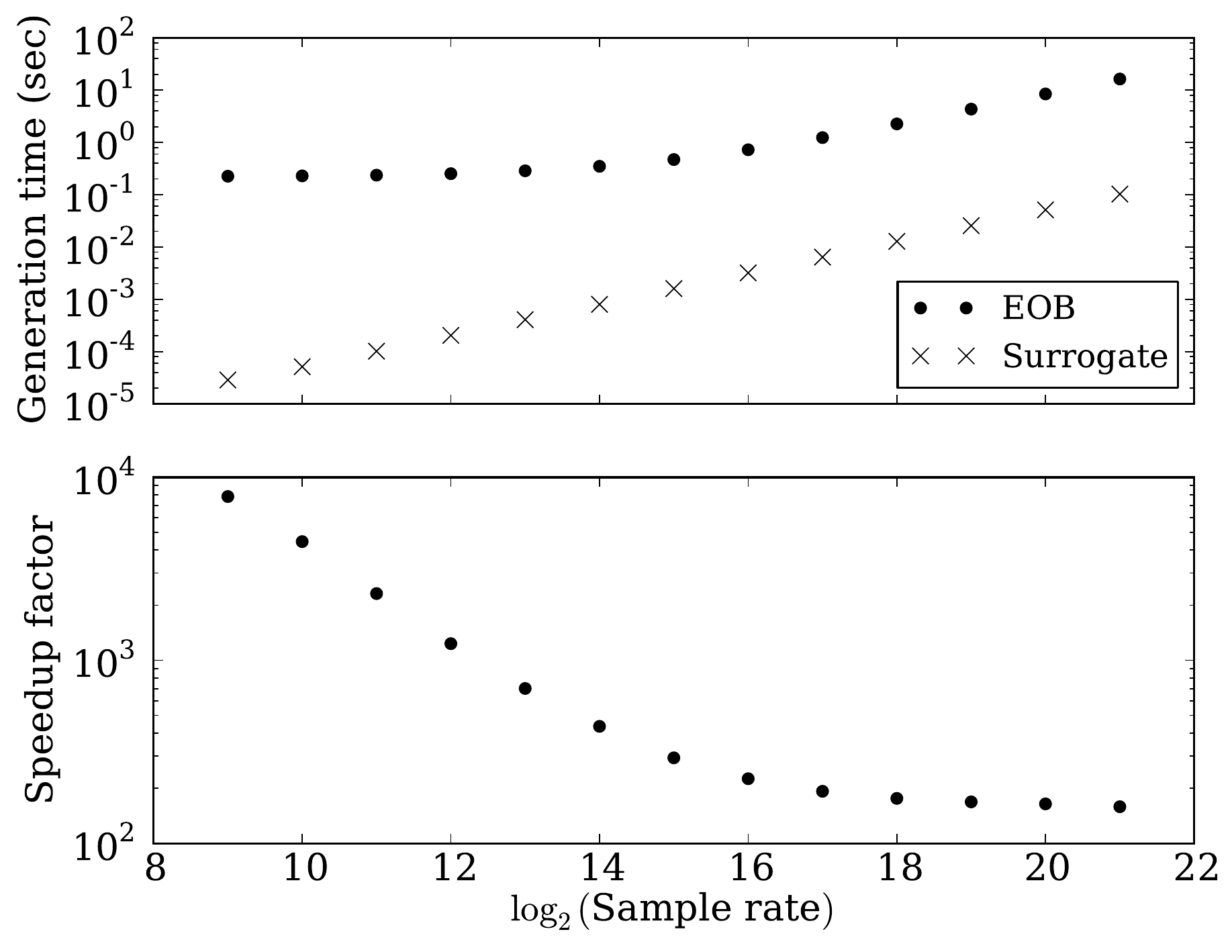}
\caption{{\bf Top}: Average time to generate a single fiducial EOB waveform from a standard EOB code (circles) and through evaluation of its surrogate (crosses).
Here we show results for the nominal example when using polynomial least squares fits for the amplitudes and phases. 
{\bf Bottom}: The speedup, defined as the ratio of waveform generation times for EOB-LAL code 
to the surrogate model. 
}
\label{fig:EOB_ExampleTiming}
\end{figure}

Figure \ref{fig:EOB_ExampleTiming} shows timing results for the nominal EOB test case with $m = 10$ and a surrogate error (\ref{eq:l2_surrogate_error_d}) uniformly below $10^{-7}$ for all mass ratios between $1$ and $2$. The top panel confirms that the cost of evaluating the surrogate model is linear in the number of samples $L$, as discussed above. 

Depending on the sampling rate, the speedup in evaluating the surrogate model compared to generating an EOB waveform with the EOB-LAL code
is between two and almost four orders of magnitude. For a sampling rate of $2^{11}= 2,\!048$ Hz, which is the rate used in the S$5$ and S$6$ searches for gravitational waves from binary black holes by the LIGO-VIRGO-GEO600 collaboration \cite{Abadie:2011kd, Aasi:2012rja}, the speedup is $\approx 2,\!300$ as shown in the bottom panel of Fig.~\ref{fig:EOB_ExampleTiming}. This is about {\em three orders of magnitude faster} than the EOB-LAL code. 

The speedups indicated here are not an artifact of studying waveforms from binaries with nearly equal masses. Repeating these experiments for waveforms with mass ratios 
from $9$ to $10$ (chosen so that the typical duration $\approx 11,\!000M$ and number of waveform cycles $\approx 80$ are comparable to our nominal EOB example), we find that only $m = 15$ reduced basis waveforms are needed to span the space with $\sigma_m = 10^{-11}$. The resulting surrogate model has an error from \eqref{eq:l2_surrogate_error_d} of $\lesssim 8 \times 10^{-9}$ with a corresponding speedup in the online stage of about $5,\!000$ at a sampling rate of $2,\!048$ Hz. Again, the speedup is about three orders of magnitude.

\begin{figure}[ht]
\includegraphics[width=0.98\linewidth]{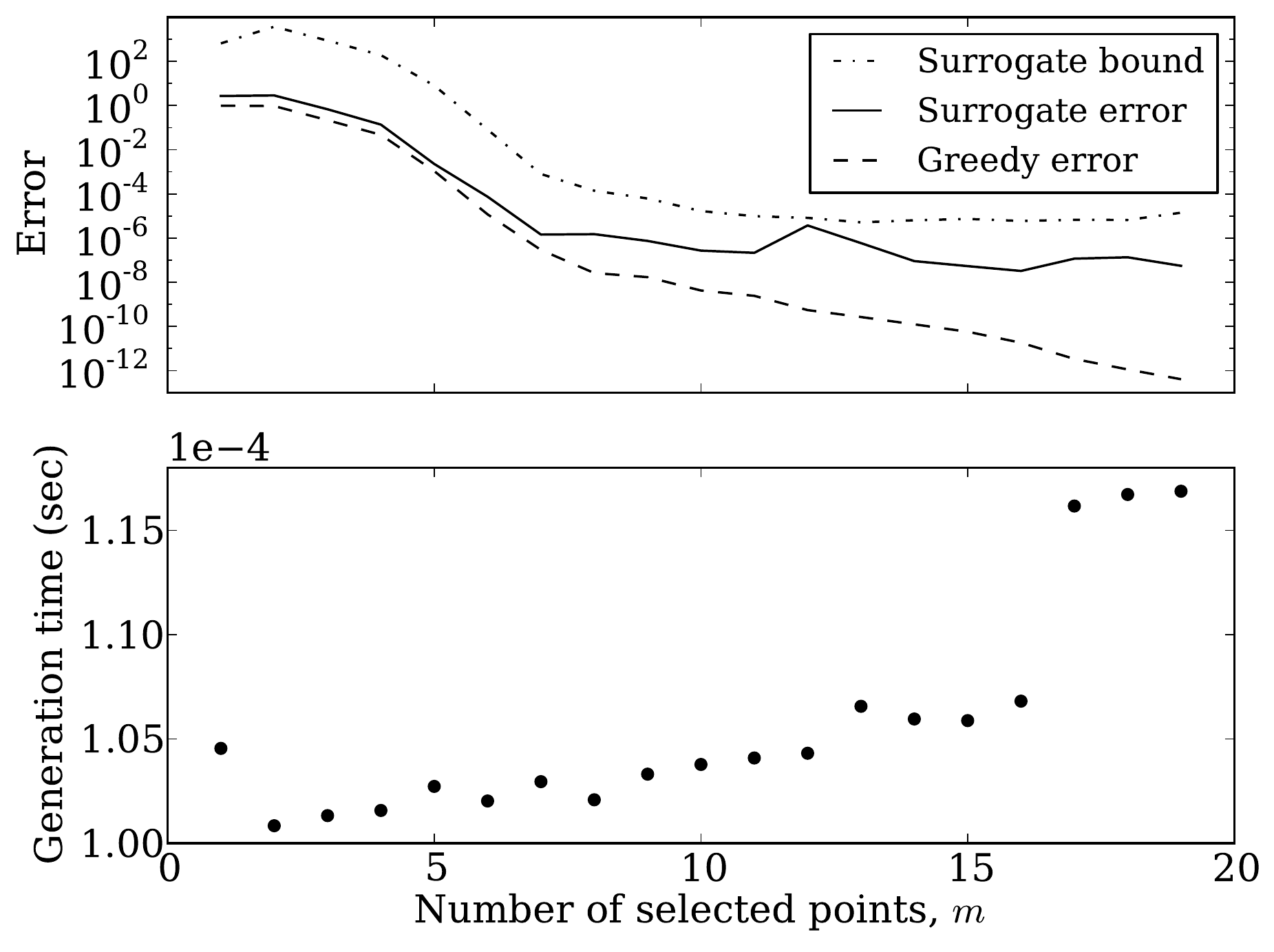}
\caption{{\bf Top}: The greedy error in \eqref{eq:greedyErr} computed for $1,\!000$ randomly selected waveforms (dashed) and the error \eqref{eq:l2_surrogate_error_d} of the resulting surrogate model (solid) as a function of the number of basis waveforms $m$. Due to fitting errors (see Sec.~\ref{sec:LS}) the surrogate error is roughly constant after $m=7$ implying little practical gain in using more than $7$ basis waveforms. The dash-dotted line shows an averaged error bound provided by the right side of (\ref{eq:sErrorb}).
{\bf Bottom}: Average time to generate a surrogate waveform (at a sampling rate of $2,\!048$ Hz) as a function of $m$. As expected there is only mild growth with $m$. 
}
\label{fig:EOB_ExampleVsM}
\end{figure}

As already mentioned in Sec.~\ref{sec:assess_surrogate}, the fitting step for building the surrogate potentially introduces the largest errors in the surrogate model. For the EOB example, these largest errors are still small (see Fig.~\ref{fig:EOB_ExampleSurrogateError}) and suggest that one does not need to include all $19$ basis waveforms/empirical times in order to yield a sufficiently accurate approximation. The top panel of Fig.~\ref{fig:EOB_ExampleVsM} shows the surrogate error in (\ref{eq:l2_surrogate_error_d}), maximized over $1,\!000$ randomly selected waveforms, as a function of the number of selected RB waveforms $m$. After $m=7$ there is little
to be gained by including more basis waveforms because the surrogate error is roughly constant until $m=19$ while, from the bottom panel of Fig.~\ref{fig:EOB_ExampleVsM}, its evaluation time continues to grow with $m$. The dash-dotted line in the top panel shows the expected error computed by averaging the surrogate's error bound (\ref{eq:sErrorb}) over $q$. Taking the average (maximum) of (\ref{eq:sErrorb}) over $q$ we are guaranteed surrogate errors of better than $10^{-5}$ ($5 \times 10^{-5}$), which is sufficient for many GW applications. The actual errors, which might be inaccessible for some fiducial waveform models, are better than $10^{-7}$ (c.f.~Fig.~\ref{fig:EOB_ExampleSurrogateError} and the solid curve in the top panel of Fig.~\ref{fig:EOB_ExampleVsM}).

\section{Astrophysical surrogates}
\label{sec:astro}

For pedagogical considerations we have primarily focused on the $(2,2)$ mode of non-spinning EOB waveforms in the range $q \in [1,2]$ and about $12,\!000M$ in duration. In this section we build surrogate models for a variety of astrophysical sources relevant for detection templates and parameter estimation with gravitational wave detectors. Typically six or fewer digits of accuracy suffice for these applications. We therefore build surrogates here with this criteria in mind by considering less ambitious error requirements of $\approx 10^{-6}$ instead of $\approx 10^{-9}$. The surrogate models presented here also have more cycles, cover larger mass ratio intervals, and include higher spherical harmonics. Surrogates built for these more challenging scenarios continue to be both accurate and fast to evaluate. Most  importantly, we can apply {\em exactly} the same method described earlier in Sec.~\ref{sec:SurrogateModels}.  

Table \ref{tab1} provides a summary of the surrogate models presented here, which we discuss in more detail below. Surrogates $1$ and $2$ were discussed earlier in Sec.~\ref{sec:SurrogateModels}. In Sec.~\ref{sec:SurrogateModels} we show that the time-domain overlap error (i.e.~the mismatch) is one-half the $L_2$ error measure we use in Table \ref{tab1}. Since our goal is to directly match the output of the EOB-LAL code we do {\em not} minimize over intrinsic or extrinsic parameters to compute the error. Hence, both the faithfulness and effectualness diagnostics will be even smaller than those implied by Table \ref{tab1}. 

A surrogate model needs to be computed ``once and for all time" for a given set of specifications and so we always strive to make surrogates with the highest possible accuracy unless otherwise indicated. By working to high accuracies one has guaranteed results equivalent to using the full underlying model (in this case EOB waveforms) without the need for special case-by-case studies of systematic biases. For particular applications reduced accuracy could be acceptable, especially at the benefit of faster model evaluations. We shall not pursue such application specific optimizations here.

\begin{table*}
\begin{tabular}{c || c | c | c | c | c | c | c | c |}
Case    &~ $q$ Interval   ~&~ Duration ($M$) ~&~ Mode  ~& $m$  & $L_2$ error ~ &~  $L_{\rm inf}$ error  ~& ~ Evaluation (sec) ~ &~ Speedup ~ \\
\hline
$1$ (A)  & $[1,2]$   & $12,\!240$   & $(2,2)$         & $19$ & $6\times10^{-8}$ & $3\times10^{-3}$ & $1\times10^{-4}$ & $1,\!900$ \\
$1$ (B)  & $[1,2]$   & $12,\!240$   & $(2,2)$         & $10$ & $3\times10^{-7}$ & $4\times10^{-3}$ & $1\times10^{-4}$ & $2,\!300$ \\
$2$      & $[9,10]$  & $11,\!103$   & $(2,2)$         & $15$ & $1\times10^{-7}$ & $2\times10^{-3}$ & $1\times10^{-4}$ & $5,\!000$ \\
$3$      & $[1,4]$   & $12,\!240$   & $(2,2)$         & $25$ & $2\times10^{-7}$ & $2\times10^{-3}$ & $1\times10^{-4}$ & $1,\!800$ \\
$4$      & $[1,6]$   & $12,\!240$   & $(2,2)$         & $25$ & $7\times10^{-8}$ & $3\times10^{-3}$ & $1\times10^{-4}$ & $1,\!900$ \\
$5$      & $[1,8]$   & $12,\!240$   & $(2,2)$         & $35$ & $6\times10^{-8}$ & $3\times10^{-3}$ & $2\times10^{-4}$ & $1,\!700$ \\
$6$      & $[1,10]$  & $12,\!240$   & $(2,2)$         & $40$ & $6\times10^{-8}$ & $2\times10^{-3}$ & $2\times10^{-4}$ & $1,\!700$ \\
$7$      & $[1,2]$   & $80,\!750$   & $(2,2)$         & $30$ & $7\times10^{-7}$ & $2\times10^{-2}$ & $5\times10^{-4}$ & $1,\!000$ \\
$8$ (A)  & $[1,2]$   & $191,\!840$  & $(2,2)$         & $20$ & $1\times10^{-3}$ & $3\times10^{-2}$ & $7\times10^{-4}$ & $1,\!100$ \\
$8$ (B)  & $[1,2]$   & $191,\!840$  & $(2,2)$         & $35$ & $1\times10^{-6}$ & $2\times10^{-2}$ & $1\times10^{-3}$ & $750$ \\
$9$      & $[1,2]$   & $12,\!240$   & $(2,1)$         & $15$ & $8\times10^{-7}$ & $1\times10^{-2}$ & $1\times10^{-4}$ & $2,\!100$ \\
$10$     & $[1,2]$   & $12,\!240$   & $(3,3)$         & $15$ & $5\times10^{-6}$ & $3\times10^{-2}$ & $1\times10^{-4}$ & $2,\!300$ \\
$11$     & $[1,2]$   & $12,\!240$   & $(4,4)$         & $15$ & $2\times10^{-5}$ & $4\times10^{-2}$ & $1\times10^{-4}$ & $2,\!100$ \\
$12$     & $[1,2]$   & $12,\!240$   & $(5,5)$         & $15$ & $1\times10^{-5}$ & $5\times10^{-2}$ & $1\times10^{-4}$ & $2,\!200$ \\
\end{tabular}
\caption{Errors, evaluation times, and speedups of surrogate models for various intervals of mass ratios, durations in time (i.e., number of cycles), and spin-weighted spherical harmonic modes.
\label{tab1}
}
\end{table*}

\begin{figure}[ht]
\includegraphics[width=\columnwidth]{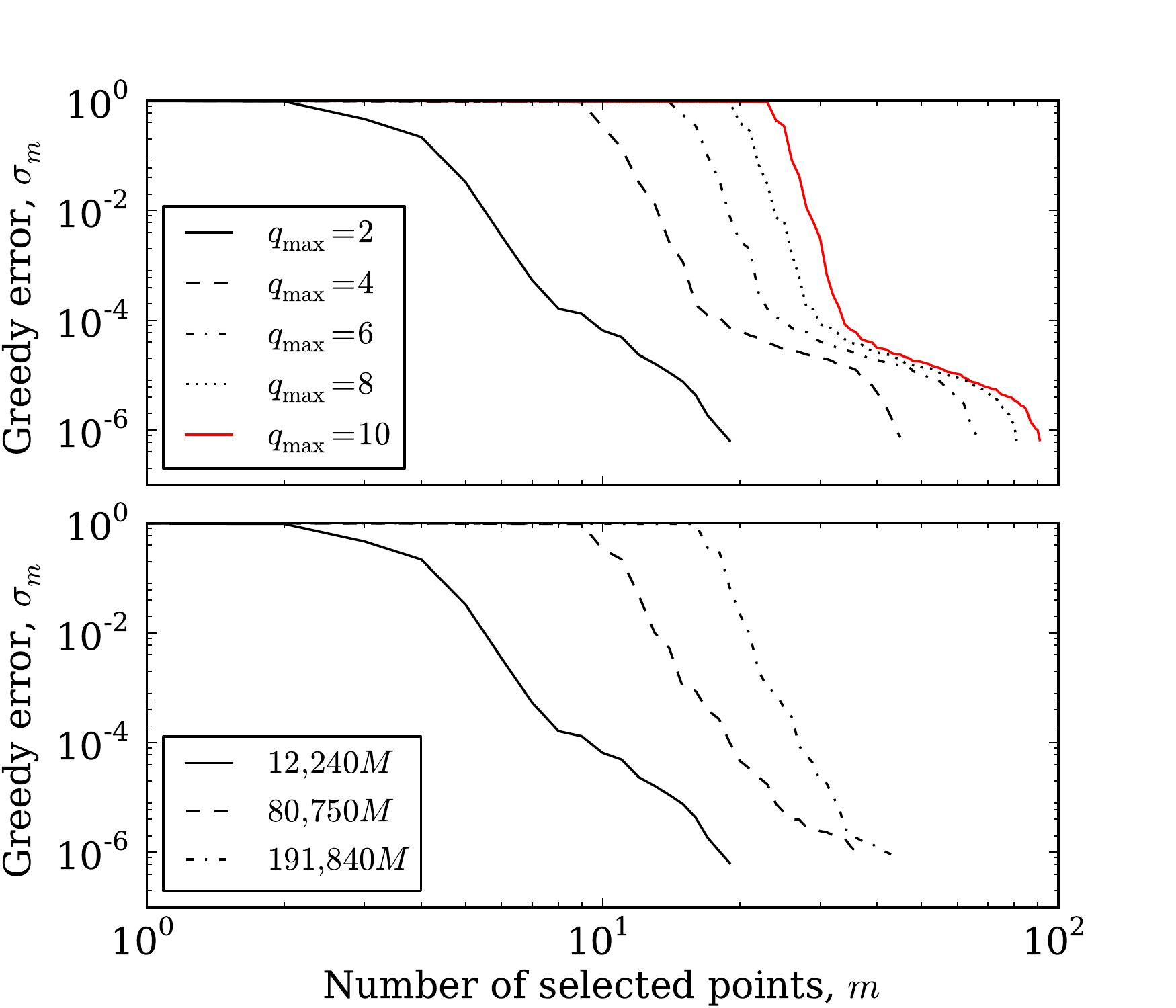}
\caption{
{\bf Top}: The greedy error in (\ref{eq:greedyErr}) computed for waveforms of fixed duration and mass ratios in $[1, q_{\rm max}]$ with $q_{\rm max} = 2$, $4$, $6$, $8$ and $10$. The curves correspond to cases $1$, $3$, $4$, $5$ and $6$ from Table \ref{tab1}. 
{\bf Bottom}: Greedy error for mass ratios in $[1,2]$ with different durations in time, thus numbers of cycles. The curves correspond to cases $1$, $7$, and $8$ from Table \ref{tab1}.
} 
\label{fig:q_t_surrogatescaling}
\end{figure}

\subsection{Larger mass ratios (Cases $1$-$6$)}

The top panel in Fig.~\ref{fig:q_t_surrogatescaling} shows that the number of reduced basis waveforms needed to approximate larger mass ratios accurately increases only mildly. In particular, the number of basis functions to achieve $6\times10^{-8}$ accuracies grows from $19$ to $40$ when $q_{\rm min} = 1$ and $q_{\rm max}$ is raised from $2$ to $10$. Furthermore, surrogates built for the intervals $[9,10]$ and $[1,2]$ use a total of $33$ basis functions, which is nearly the amount ($40$) needed for the entire $[1,10]$ range. This feature is typical of global approximation methods (such as reduced basis and empirical interpolation) since they tend to promote sparseness whenever the underlying model is sufficiently smooth.

\subsection{Longer durations (Cases $1$, $7$ and $8$)}

The bottom panel in Fig.~\ref{fig:q_t_surrogatescaling} shows the number of RB functions needed to accurately cover waveforms with longer durations (i.e., more cycles) also increases mildly. For the longest, most accurate surrogate, $8$B, the evaluation time is as large as $10^{-3}$ seconds, which is an order of magnitude larger than the shorter, but otherwise equivalent, case in surrogate $1$. However, the EOB-LAL code also runs slower. Thus, the overall speedup is found to be about $750$. The lower accuracies required for gravitational wave detection templates (as opposed to the higher accuracy standards for parameter estimation) imply that the speedup can be improved to about $1,\!100$ (see case $8$A).

\subsection{Higher harmonics (Cases $1$ and $9$-$12$)}

Gravitational waveforms have multiple spin-weighted spherical harmonic modes. Surrogate models must accommodate these multi-mode functions in order to maximize their usefulness. One direct approach is to build a surrogate for each mode separately using exactly the same steps described earlier in Sec.~\ref{sec:SurrogateModels}. The resulting multi-mode surrogate model is then defined by the set of single-mode surrogates. Some $q=1$ modes are identically zero, and while the reduced basis can exactly approximate zero modes they  slightly complicate the treatment of parametric fits (e.g.,~the phase is undefined): we construct fits on an open interval $q \in (1,2]$. Our surrogates are thus defined on this open interval with $q=1$ modes given by zero. To asses the error of each multi-mode surrogate model we continue to draw $1,\!000$ random waveform samples from the closed interval $[1,2]$.

As a demonstration, we have built surrogate models for the $(2,1)$, $(2,2)$, $(3,3)$, $(4,4)$ and $(5,5)$ modes in the same physical and parametric ranges used for the nominal EOB example problem (surrogate $1$) considered throughout this paper. These five modes exhaust the currently known ones provided by the EOB model. Compared to the $(2,2)$ mode, surrogates built for these higher harmonics are more sensitive to peak alignment (see App.~\ref{sec:EOB}) which translates into larger surrogate errors. These errors are still small and, furthermore, higher harmonics typically have less contribution to the overall gravitational wave strain measurement.

Another way to build multi-mode surrogate models starts by integrating the complex scalar product in (\ref{eq:dotproduct}) over the angles $(\theta, \phi)$ on the $2$-sphere. The orthonormality of the spin-weighted spherical harmonics implies a sum over the scalar product in (\ref{eq:dotproduct}) for each mode,
\begin{align}
	\langle h(\cdot; \mlam_1), h(\cdot; \mlam_2) \rangle &  = \! \int_{S^2} {\hskip-0.1in} d\Omega {\hskip-0.025in} \int _{t_{\rm min}} ^{t_{\rm max}} {\hskip-0.2in} dt \, h^*(t, \theta, \phi; \mlam_1) h (t, \theta, \phi; \mlam_2) \nonumber \\
		& = \sum_{\ell, m} \int _{t_{\rm min}} ^{t_{\rm max}} {\hskip-0.2in} dt \, h_{\ell m}^*(t; \mlam_1) h_{\ell m} (t; \mlam_2)
\label{eq:newdotproduct}
\end{align}
which is used for constructing the norm in the greedy error in (\ref{eq:greedyErr}). When the RB-greedy algorithm is performed using the scalar product in (\ref{eq:newdotproduct}), {\it all modes} contribute to selecting the relevant parameters in the space. Likewise, the residual used for finding the empirical nodes in Step 6 of Algorithm \ref{alg:EIM}, and other quantities required by the EIM, can also be integrated over the angles of the $2$-sphere so that all modes can contribute to building the global empirical interpolant. 

We will not cover here all possible variations for multi-mode waveforms since the choices made are largely dependent upon the specific application and waveform model studied. This is typical in problems that require learning from data, such as this one.

\subsection{Building costs}

Ignoring training set generation, the surrogates listed in Table \ref{tab1} typically took between $5$ and $20$ minutes to build. However, we have found the main cost, both in terms of computational and memory requirements, to be in creating the training set. These costs are significantly greater than what might be expected from any particular surrogate's properties (e.g.~sampling rate and duration). As discussed further in App.~\ref{sec:EOB}, the error of an EOB surrogate is dominated by the error in resolving and localizing the waveform's peak. Consequently, we must generate training data having well-resolved waveform peaks. For example, the EOB-LAL code was called using a sampling rate of $2^{20}$ Hz to generate the training set data for cases $1$ through $6$ in Table \ref{tab1}. Surrogate $6$ was trained on $2001$ EOB waveforms, which took nearly $8$ hours to generate. Waveform generation times quoted throughout this paper, for both the EOB-LAL code and surrogate evaluations, do not depend on these settings in anyway whatsoever. 

For longer waveforms, such as cases $7$ and $8$, we are unable to maintain these high sampling rates. A single waveform cannot be produced (on a personal computer) due to the larger memory overhead. In lieu of using higher memory nodes we instead decreased the sampling rate to $2^{18}$ Hz to make the problem more manageable. As a result, the pointwise (maximum) waveform errors increase but remain acceptably small in many cases.

\section{Concluding remarks and outlook} 
\label{sec:summary}

We introduced a solution to the problem of quickly and accurately generating predictions for a given family of gravitational waveforms. The solution constructs a surrogate for this fiducial set of waveforms in three offline steps. In the first step, a reduced basis is generated that spans the space of waveforms in the given range of parameters. In the second step, an application-specific (i.e., empirical) interpolant is constructed using only these $m$ reduced basis waveforms. The empirical interpolation method selects a corresponding set of $m$ times that are used to build the interpolant but requires knowing the fiducial waveform at any parameter value at those times in order to evaluate the interpolant. In the third step, we complete the offline part by implementing a fit for the parametric dependence of the waveform's phase and amplitude at each empirical time. In this way, the value of the fiducial waveform at each empirical time can be estimated and then fed into the empirical interpolant. The result of these three offline stages is an accurate surrogate model (\ref{eq:surrogate100}) for the underlying family of waveforms that is cheap to evaluate for any parameter value in the considered range. 

Surrogate models offer a new and complementary approach to other modeling endeavors. Indeed, our goal is to clone the input-output functionality of an existing waveform generation code thereby permitting fast evaluations for any task of interest. Consequently, we have intentionally worked in a detector-independent context, to very high accuracies, and without regard for the systematic errors of the underlying waveform family. To ensure the surrogate can be used in place of an underlying model (without introducing bias) it is best to be as accurate as possible. However, for particular applications one may wish to sacrifice accuracy at the benefit of even faster surrogate model evaluations.

The standard paradigm for fast online evaluation of new solutions within reduced order modeling frameworks (see, e.g., \cite{Quarteroni} for a review)  
is to numerically solve a small problem that is essentially a projection of the original problem onto the basis built in the offline stage. Nonlinear terms or non-affinely parametrized problems can be dealt with using the EIM \cite{chaturantabut:2737}. This approach has some advantages. For example, for many problems of interest rigorous error bounds can be guaranteed for the resulting output, which is often referred to as a {\it certified} approach. 

In this paper we deviated from this standard course and sought a different and more heuristic one for two major reasons specific to gravitational waveforms. First, the complexity of projecting the full nonlinear Einstein equations onto a basis to obtain a certified approach is highly nontrivial. Second, our goal has been to develop a {\em non-intrusive} approach that does not resort to manipulating, in any way, the original equations and codes that generate the fiducial waveform model.
Of course, such equations have to be used to generate the fiducial waveforms in the offline stage in order to build the reduced basis to start the construction of the surrogate model. However, the approach introduced in this paper does not intrude upon or require editing those codes.

In order to demonstrate the basic ideas and methods in this paper, we have focused on surrogate models for single-mode, non-spinning black hole binary EOB waveforms \footnote{Recently, while this paper was under publication review, surrogate models for non-precessing EOB waveforms have been built in \cite{Purrer:2014fza}.}. For mass ratios in $[1,2]$, we find that evaluating the surrogate is three orders of magnitude faster than generating EOB waveforms in the standard way. 
However, the construction of the surrogate model is not limited to such a short range of mass ratios, to non-spinning binaries, nor to single-mode waveforms. We demonstrated this in Sec.~\ref{sec:astro} by building surrogates for astrophysically motivated problems relevant for template bank generation and parameter estimation studies with gravitational wave detectors. Regarding the range of mass ratios (or other parameters), depending on the application and the target accuracy, a partitioning of the parameter space might provide faster online queries. This issue is familiar when solving differential equations where one may choose to use a single domain or utilize a domain decomposition, as with a spectral or hp-element approach (see, e.g., \cite{karniadakis2005spectral, Hesthaven:2002:NHM:637611.637619}). Similar tools for parameter space subdomain decomposition, known as {\it hp-greedy} algorithms, have been employed as an adaptive sampling strategy for large problems (see \cite{Fares20115532, Eftang2011, Eftang:2012:TCR:2197457.2197477, NME:NME3327} for further details). 

A preliminary cost-benefit analysis of domain decomposition is provided by Table \ref{tab1}, which summarizes all surrogates considered in this paper. Taking the $19$ basis functions for the $[1,2]$ range (surrogate $1$ in Table \ref{tab1}) as indicative of the reduced basis size needed for each successive integer range of mass ratios up to $q=10$, a naive scaling with a domain decomposition approach would suggest $\approx 19 \times 9 = 171$ basis elements. Compare this number to the $40$ elements needed for the whole range $[1,10]$. While the latter gives fewer basis functions for representing EOB waveforms across the whole $[1,10]$ range, the cost to evaluate surrogate models increases with $m$. For example, if one were interested in only waveforms with $q$ from $9$ to $10$ then surrogate $2$ in Table \ref{tab1} would be preferable in terms of speedup since $m=15$, not $40$. Such optimizations are application specific and, as such, were not pursued in this paper.

Finally, the method presented in this paper for building a surrogate model can be applied to other waveform families, including precessing inspiral waveforms and multi-mode inspiral-merger-ringdown waveforms such as those from NR simulations of binary black hole coalescences. We anticipate extremely large speedup factors for predicting a NR waveform with a surrogate model compared to solving the Einstein equations for the same parameters because the cost of evaluating the surrogate is independent of the offline costs required to build it. Given that a single production-quality simulation for a non-spinning equal mass binary takes around $\sim 10^4-10^5$ hours and predicting a single-mode waveform with a NR-based surrogate model takes about $10^{-4}$ seconds (as implied by Fig.~\ref{fig:EOB_ExampleVsM}), it follows that one may expect speedup factors of 
$\sim 10^{11}$ or more.

\section{Acknowledgments} 
We thank Frank Herrmann and Evan Ochsner for help during this project, including some software tools, as well as Yi Pan, Alessandra Buonnano, and Collin Capano for helpful discussions about the EOB model and its generation using the LAL code. 
This work was supported in part by NSF grants PHY-1208861 and PHY-1005632 to the University of Maryland and by NSF grant PHY-1068881 and CAREER grant PHY-0956189 to the California Institute of Technology. 

\appendix

\section{The reduced basis method} \label{sec:RBapp}

We use a greedy algorithm to build a reduced basis (RB), which accurately approximates any fiducial waveform within the given parameter ranges (see, e.g., Ref.\,\cite{Field:2011mf}). The greedy algorithm, outlined in Algorithm \ref{alg:Greedy}, takes as inputs a discretization of the parameter space ${\cal T} \equiv \{\mlam_i\}_{i=1}^M$ (or the training space) and the associated waveforms, an arbitrary parameter $\mLam_1 \in {\cal T}$ (or seed), and a threshold error $\epsilon$ for a target representation accuracy (or greedy error). The output consists of the $m$ RB waveforms and $m$ greedy points.

\hspace{0.5cm}

{\scriptsize
\begin{algorithm}[H]
\caption{Greedy algorithm for reduced basis}
\label{alg:Greedy}
\begin{algorithmic}[1]
\State {\bf Input:} $ \{ \mlam_i \, , h(\cdot; \mlam_i) \}_{i=1}^M$,  $\epsilon$ 
\vskip 10pt
\State Set $i=0$ and define $\sigma_0 = 1$
\State {\bf Seed choice} (arbitrary):  $\mLam_1 \in {\cal T}$, $e_1 = h(\cdot; \mLam_1)$
\State RB = $\{ e_1 \}$ 
\While{$\sigma_i \ge \epsilon$}
\State $i=i+1$
\State $\sigma_i = \max_{ \mlam \in {\cal T} } \| h(\cdot; \mlam ) - {\cal P}_{i} h(\cdot; \mlam ) \|^2$ 
\State $\mLam_{i+1} = \text{argmax}_{ \mlam \in {\cal T} } \| h(\cdot; \mlam ) - {\cal P}_{i} h(\cdot; \mlam ) \|^2 $  
\State $e_{i+1} = h(\cdot; \mLam_{i+1} ) - {\cal P}_{i} h(\cdot; \mLam_{i+1} )$ ~(Gram-Schmidt)
\State $e_{i+1} = e_{i+1} / \| e_{i+1} \|$ {\hskip0.725in} (normalization)
\State RB = RB $\cup \, e_{i+1}$
\EndWhile
\vskip 10pt
\State {\bf Output:} RB $\{ e_i \}_{i=1}^m$ and greedy points $\{ \mLam_i \}_{i=1}^m$
\end{algorithmic}
\end{algorithm}
}

The naive implementation of the classical Gram-Schmidt procedure can lead to a numerically ill-conditioned algorithm. This is related to the fact that the Gramian matrix, which would have to be inverted, can become  nearly singular \cite{taylorLS}. To overcome this we use an iterated Gram-Schmidt algorithm or a QR decomposition in step $9$. See \cite{Ruhe1983591,Giraud} for discussions about the conditioning and numerical stability of different orthonormalization procedures.

As mentioned in Sec.~\ref{sec:Greedy}, minimization over the coefficients $\{c_i\}$ in (\ref{eq:projapprox}) is satisfied by orthogonal projection ${\cal P}_m h(t;\mlam)$ of $h(t;\mb{\lambda})$ onto the span of the basis.
For example, for an orthonormal basis 
\begin{align}
& {\hskip-.5in} \left \| h (\cdot; \mlam) - \sum_{i=1}^m c_i (\mlam) e_i (\cdot) \right \|^2 \nonumber \\ 
& = \big\| h (\cdot; \mlam) \big\|^2 - \sum_{i=1}^m \big| \langle h(\cdot; \mlam), e_i(\cdot) \rangle \big| ^2 \nonumber \\
& ~~~~ + \sum_{i=1}^m \big|  \langle h (\cdot; \mlam), e_i (\cdot) \rangle - c_i (\mlam)\big|^2  \, , 
\end{align}
which takes its global minimum when 
\begin{align} 
c_i (\mlam) = \langle h(\cdot;\mlam) , e_i(\cdot) \rangle \, .
\end{align}

After applying the greedy algorithm to build a reduced basis and find the greedy points, we check that the basis accurately approximates the continuum space of waveforms for the given parameter range by verifying at a randomly chosen set of test points. 

\section{The empirical interpolation method} \label{sec:EIMapp}

The Empirical Interpolation Method (EIM) provides a sparse subset of empirical time (or frequency) nodes from which it is possible to reconstruct the waveform at any other time with very high accuracy using an application-specific interpolant. The selection of the empirical time nodes and the construction of the empirical interpolant proceeds using a greedy algorithm, which is hierarchical and is applicable to unstructured meshes in several dimensions.

Consider a basis $\{ e_i \}_{i=1}^m$ (e.g., a RB) whose span approximates the functions of interest. Let $\{t_i\}_{i=1}^L$ denote a set of $L$ time samples and define the $L$-vector $\vec{t} = \left(t_1, t_2, \dots, t_L \right)^{\dagger }$. For compactness of notation, denote other functions evaluated at these time samples as vectors so that, for example,  $\vec{h}(\mlam) := h(\vec{t}; \mlam)$ and $\vec{e}_i := e_i(\vec{t})$. 

Given an input of $m$ evaluated basis functions  
$\{ \vec{e}_i \}_{i=1}^m$ the output of the EIM algorithm is a set of $m$ empirical nodes 
\begin{align}
\{ T_i \}_{i=1}^m \subset \{t_i\}_{i=1}^L \label{eq:EIM_points}
\end{align}
selected as a subset of $\{t_i\}_{i=1}^L$. The empirical interpolant is constructed in step $5$ of Algorithm \ref{alg:EIM}. At the $j^{\rm th}$ iteration the empirical interpolant is built from the first $j$ basis functions and nodes, 
\be
{\cal I}_j [h](t;\mb{\lambda}) = \sum_{i=1}^j C_i (\mlam) e_i(t)\, ,  \label{eq:interpdefI}
\ee
where the $C_i$ coefficients are solutions to the $j$-point interpolation problem 
\begin{align}
\label{eq:interp1000}
{\cal I}_j [h](T_k;\mb{\lambda}) = h(T_k;\mb{\lambda}) 
\end{align}
for all $\mlam$ and where $k=1, \dots, j$.

\hspace{0.5cm}

{\scriptsize
\begin{algorithm}[H]
\caption{The Empirical Interpolation Method}
\label{alg:EIM}
\begin{algorithmic}[1]
\State {\bf Input:} $\{ \vec{e}_i \}_{i=1}^m$, $\{t_i\}_{i=1}^L$
\vskip 10pt
\State $i = \text{argmax} | \vec{e}_1 |$ (\text{argmax} returns the largest entry of its argument). 
\State Set $T_1 = t_i$
\For{$j = 2 \to m$} 
\State Build ${\cal I}_{j-1} [e_j](\vec{t})$ from (\ref{eq:interpdefI}) and (\ref{eq:interp1000})
\State $\vec{r} = {\cal I}_{j-1} [e_j](\vec{t}) - \vec{e}_j$
\State $i = \text{argmax} |\vec{r}|$
\State $T_j = t_i$
\EndFor
\vskip 10pt
\State {\bf Output:} EIM nodes $\{ T_i \}_{i=1}^m$ and interpolant $\cI_m$
\end{algorithmic}
\end{algorithm}
}

Let us define a discrete norm 
\be
 \| h \|_{\rm d} = \Delta t \sum_{i=1}^L h^*(t_i) h(t_i)  \, , 
\ee
for $L$ equally spaced time samples. The empirical interpolant's error is then directly related to the greedy error \eqref{eq:greedyErr} through \cite{antil2012two} 
\begin{align}
\| h  - {\cal I}_m [h] \|_{\rm d}^2 & = \| ( \mathbb{I} - {\cal I}_m ) ( h - {\cal P}_m h )\|_{\rm d}^2 \nonumber \\
& \leq  \| (\mathbb{I} - {\cal I}_m) \|_{\rm d}^2  \| h- {\cal P}_m h \|_{\rm d}^2 \nonumber \\
& = \| {\cal I}_m \|_{\rm d}^2  \| h - {\cal P}_m h \|_{\rm d}^2 \nonumber \\
& \leq \Lambda_m \sigma_m \, ,  \label{eq:bound352}
\end{align}
where the first equality follows from ${\cal I}_m[{\cal P}_m h] = {\cal P}_m h$, $\mathbb{I}$ is the identity matrix, $\| (\mathbb{I} - {\cal I}_m) \|_{\rm d}^2 = \| {\cal I}_m \|_{\rm d}^2$ holds whenever the operator norm is induced by the vector norm (as is the case here, see Refs.~\cite{JXu_LZikatanov_2003a,DBSzyld_2006a}) and
\begin{align} \label{eq:leb_AppB} 
\Lambda_m  = \| {\cal I}_m \|_{\rm d}^2 =
\max_{\| h \|_{\rm d} = 1} \| {\cal I}_m [h] \|_{\rm d}^2
\end{align}
is a computable Lebesgue-like quantity that generally changes slowly with $m$. 
For problems with smooth dependence with respect to parameter variations we can expect an exponential decay of $\sigma_m$ with $m$ and, from the left side of (\ref{eq:errorEIM}), of the EIM's error.

In practice, $\Lambda_m$ is computed from the matrix representation of $B$ from (\ref{eq:BEIM}),
\begin{align} \label{eq:ImMatrix_AppB} 
B = E V^{-1} \, ,
\end{align}
where each column of $E = [\vec{e}_1, \dots, \vec{e}_m]$ is an evaluated reduced basis function and $V$ is the interpolation matrix defined in~\eqref{eq:InterpMatrix}. The matrix operator $B$, as written above, acts on an $m$-vector $h(\vec{T};\mlam)$ whose components are evaluations of $h$ at the empirical nodes.

\section{Details of polynomial least squares}
\label{sec:polyLS}

When performing a least squares fit we must select the degree $n_{\rm LS}$ of each of the $2m$ least squares polynomials, balancing accuracy and stability of the resulting fit. For each fit there are $m$ greedy data points so $n_{\rm LS} < m$. A small value of $n_{\rm LS}$ would result in a low accuracy fit while too large of a value can exhibit Runge's phenomenon \cite{Epperson}. Furthermore, a large value of $n_{\rm  LS}$ can fit (numerical) noise thereby leading to low quality fits (this is sometimes called {\it overfitting} \cite{Dietterich:2000:EMM:648054.743935}). Reference \cite{ACohen_MADavenport_DLeviatan_2011a} provides a computable expression for the largest $n_{\rm LS}$ that avoids this phenomenon and gives an error estimate for the resulting fit.

\begin{figure}[ht]
\includegraphics[width=0.98\linewidth]{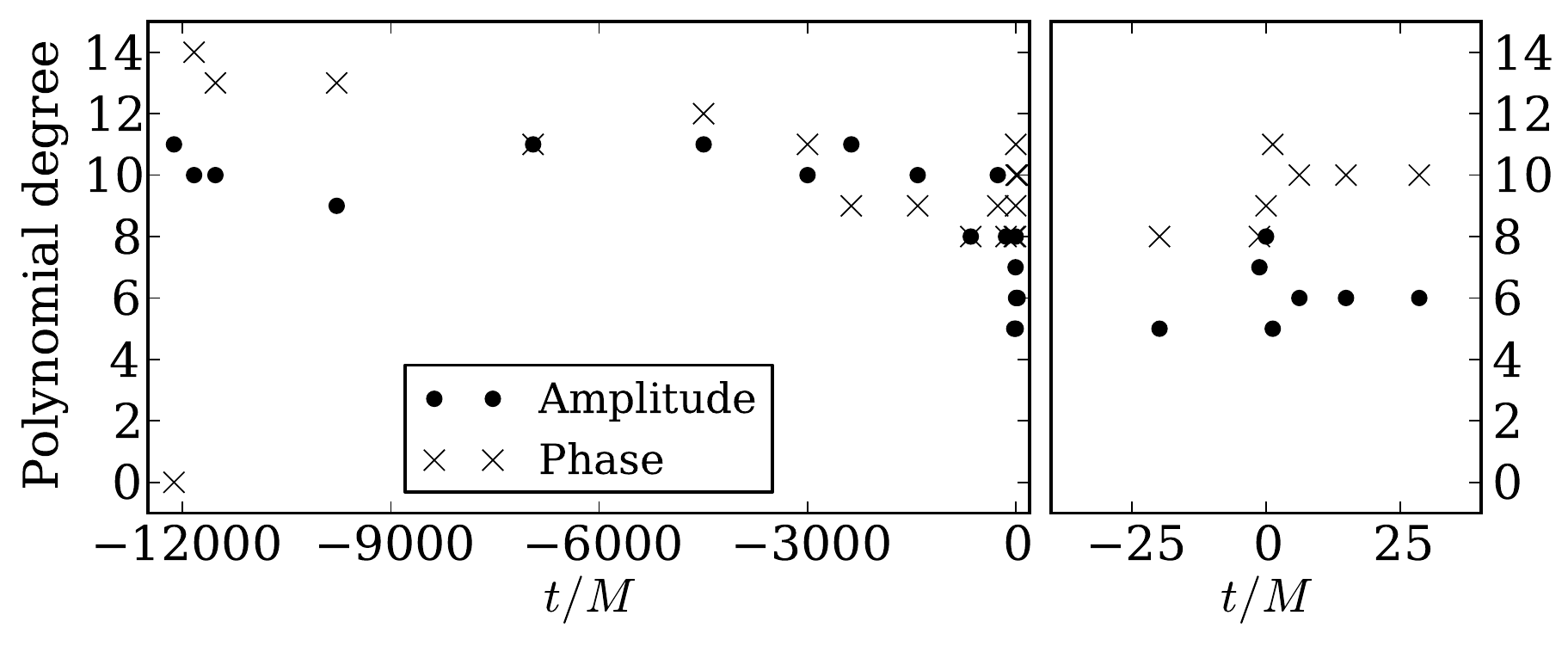}
\caption{The optimal degree of each polynomial from a least squares fit at each empirical time. Out of a possible maximum of $n_{\rm  LS}=19$, polynomial degrees between $9$ and $14$ are most often selected during the inspiral phase. The degree of the first fit for the phase is zero because the initial phases are chosen to vanish for all mass ratios.
}
\label{fig:EOB_ExamplePolyDegrees}
\end{figure}

For our nominal EOB example we proceed in a straightforward way. We construct $m$ separate fits (using only the greedy data) for all degrees $0 \leq n_{\rm  LS} < m$ and we select the one that minimizes the sum of the squared residuals relative to the training set data. This additional offline work guarantees, in a simple way, that each polynomial fit has the optimal degree. Figure \ref{fig:EOB_ExamplePolyDegrees} shows the results for our EOB test problem. We see that for empirical times in the early inspiral the optimal polynomial degrees are relatively large and decrease until merger and ringdown. This is a consequence of noisy data stemming from discrete uncertainties in locating the amplitude peak (see App.~\ref{sec:EOB} for further details).

\section{Surrogate error estimates} \label{sec:surrogateErrs}

In this appendix we derive the error bounds shown in (\ref{eq:sErrorb}) for the surrogate model.
We differentiate between the surrogate waveform model $h_{\rm S}(t; \mlam)$, whose computation requires an estimate for the waveform at each empirical node $T_i$, from the empirical interpolant ${\cal I}_m [h](t; \mlam)$, whose computation assumes the exact (fiducial) values $\vec{h}$. For any $\mlam$ we have,
\begin{align} \label{eq:surrogateErrors1}
\| {\cal I}_m [h] - h_{\rm S} \|_{\rm d}^2 & = \| {\cal I}_m [h - h_{\rm S} ] \|_{\rm d}^2 \nonumber \\ 
& \leq \Lambda_m \| h (\vec{T}) - h_{\rm S} (\vec{T})\|_{\rm d}^2 \nonumber \\
& = \Lambda_m  \Delta t \sum_{i=1}^m \left[  h(T_i) - h_{\rm S}(T_i)\right]^2 
\end{align}
with $\Lambda_m$ being the same constant defined in (\ref{eq:leb_AppB}). The first equality follows from ${\cal I}_m [h_{\rm S}(\vec{T})] = h_{\rm S}(\vec{t})$. The second line follows from the empirical interpolant's matrix representation \eqref{eq:ImMatrix_AppB}. The error in approximating an underlying model $h(t;\mlam)$ by the surrogate $h_{\rm S}(t;\mlam)$ is, for any $\mlam$,
\begin{align}
\| h_{\rm S} - h \|_{\rm d }^2  & \leq \| h_{\rm S} - {\cal I}_m [h] \|_{\rm d }^2 + \| {\cal I}_m [h] - h \|_{\rm d }^2 \nonumber \\
& \leq  \Lambda_m  \Delta t \sum_{i=1}^m \left[  h(T_i) - h_{\rm S}(T_i) \right]^2 + \Lambda_m \sigma_m \, ,\label{eq:AppsurrogateErrors2}
\end{align} 
which follows from the error bounds (\ref{eq:surrogateErrors1}) and (\ref{eq:bound352}) (or (\ref{eq:errorEIM})) as well as the triangle inequality. Notice that $\Lambda_m$ and $\sigma_m$ are computable quantities as are the differences  $h(T_i) - h_{\rm S}(T_i)$, which are only due to least square fitting errors.

\section{On generating the fiducial EOB waveform family} \label{sec:EOB} 

In this paper we demonstrated how to build a surrogate using an EOB model of non-spinning binary black hole coalescence waveforms. 
Here, we discuss some of the technical details regarding how these EOB waveforms were generated.

The specific version of the model that we used is from Ref.~\cite{Pan:2011gk} and implemented in the routine EOBNRv2 as part of the publicly available LIGO Analysis Library (LAL) Suite \footnote{git hash 59c12886b026c863397f191e6c2ca69ef3498616}. Other versions and models are equally applicable (e.g.~\cite{Damour:2012ky}). In its simplest description, the code takes as input a starting frequency $f_\mathrm{min}$ and the mass components $m_1$ and $m_2$. From initial conditions, determined through post-Newtonian expressions, the EOB differential equations are solved to give the system's orbital evolution until merger, which is defined to be the time at which the orbital frequency begins to decrease. From the compact binary system's orbit a gravitational wave is generated up to the time of merger, after which quasinormal modes are attached. 

Our nominal EOB example uses a training space of mass ratios $q\in[1,2]$. We sampled this parameter range with $501$ equally spaced points, solving the original model at each $q$ using the aforementioned code. We checked that this number of training set samples was dense enough to reach the convergent regime for building a faithful reduced basis representation.

We generated the EOB waveforms with $f_\mathrm{min} = 9$Hz and $m_1 + m_2 = 80 M_{\odot}$, which corresponds to roughly $65 - 70$ waveform cycles before merger in the $(2,2)$ mode. We avoided generating short waveforms (where the initial radial separation is less than $20M$) because the ODE initial data could become less accurate. The waveform's coalescence phase was determined implicitly through initial data instead of specifying a particular value \footnote{This can be achieved by setting the value of the EOBNRv2 variable sSub to zero.}. The relevant $(2,2)$ modes $h_+^{22} (t)$ and $h_\times^{22} (t)$, as opposed to their spin weighted values, comprised the training set. 

\begin{figure}[ht]
\includegraphics[width=0.98\linewidth]{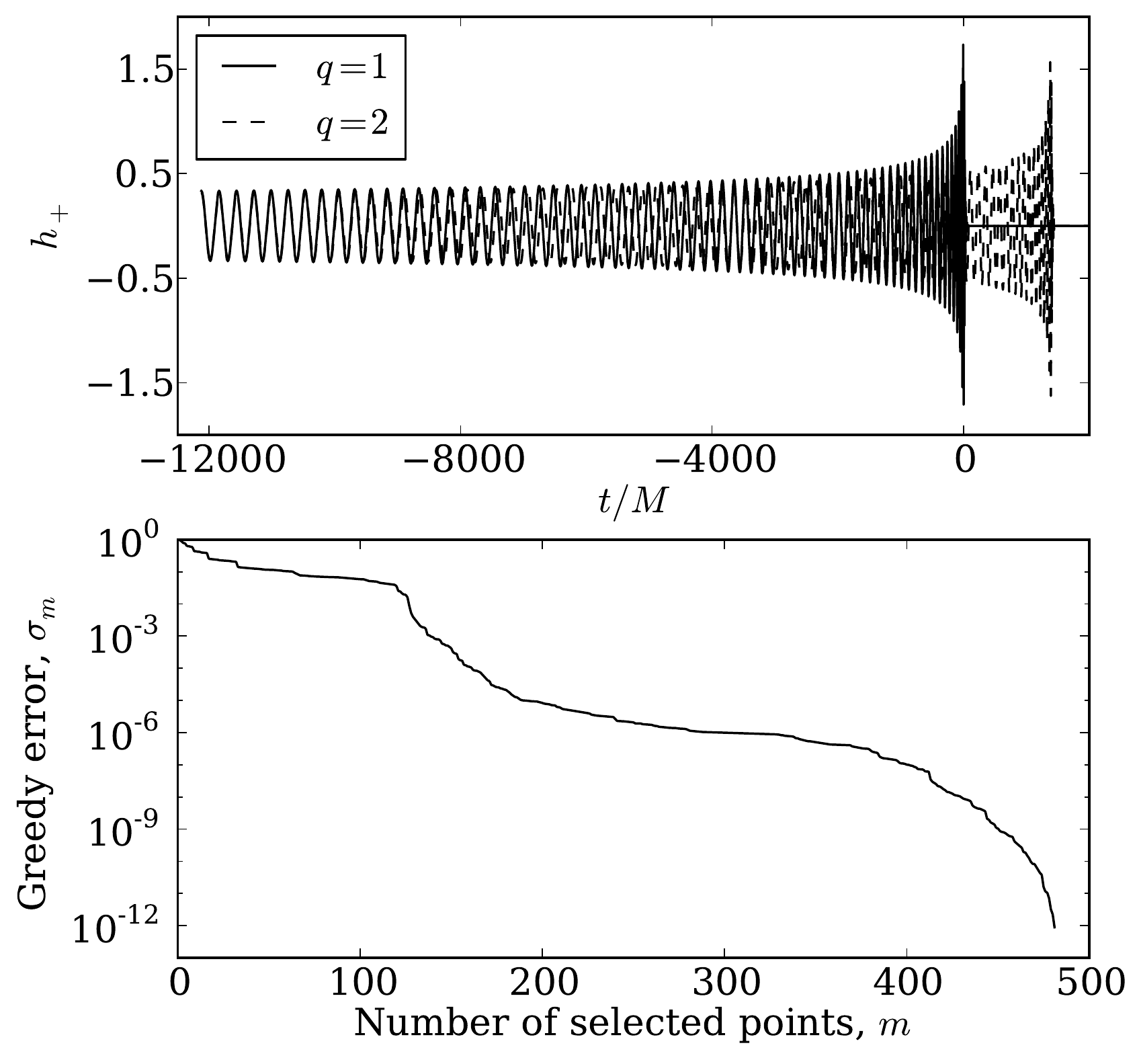}
\caption{{\bf Top}: EOB waveforms for $q=1,2$ starting at the same initial frequency but not aligned at the peak amplitudes. {\bf Bottom}: Not aligning the waveforms results in more reduced basis elements needed to accurately span the space of waveforms. Here we see that nearly all $501$ points in the training space are selected whereas only $19$ points are required if the waveforms are aligned at the peak amplitude (compare with Fig.~\ref{fig:EOB_ExampleGreedy}).} 
\label{fig:EOB_ExampleNoAlign}
\end{figure}

Waveforms generated by EOBNRv2 are automatically aligned at $f_\mathrm{min}$ and thus have lengths that depend on their mass ratio.
A typical example is shown in the top panel of Fig.~\ref{fig:EOB_ExampleNoAlign}. We have found that, when applied to a set of waveforms with varying lengths, the greedy error (\ref{eq:greedyErr}) has a very slow decay rate as indicated by the bottom panel of Fig.~\ref{fig:EOB_ExampleNoAlign}.

To overcome this, we shift each waveform in time so that their peak amplitudes are aligned. 
We first align all waveforms in the training set in this way and then ``chop off'' the beginning portions so that all waveforms have a length (from start to peak amplitude) equal to that of the shortest waveform (here, $q=1$).  
Next, we adjust each waveform's phase \eqref{eq:AmpPhase} to be initially zero. The benefits of waveform alignment are evident from the curve in Fig.~\ref{fig:EOB_ExampleGreedy}, which should be compared with the pre-alignment case shown in the bottom panel of Fig.~\ref{fig:EOB_ExampleNoAlign}. For example, to achieve a greedy error of $10^{-7}$ one needs $\approx 7$ ($400$) with (without) peak alignment.

\begin{figure}[ht]
\includegraphics[width=0.98\linewidth]{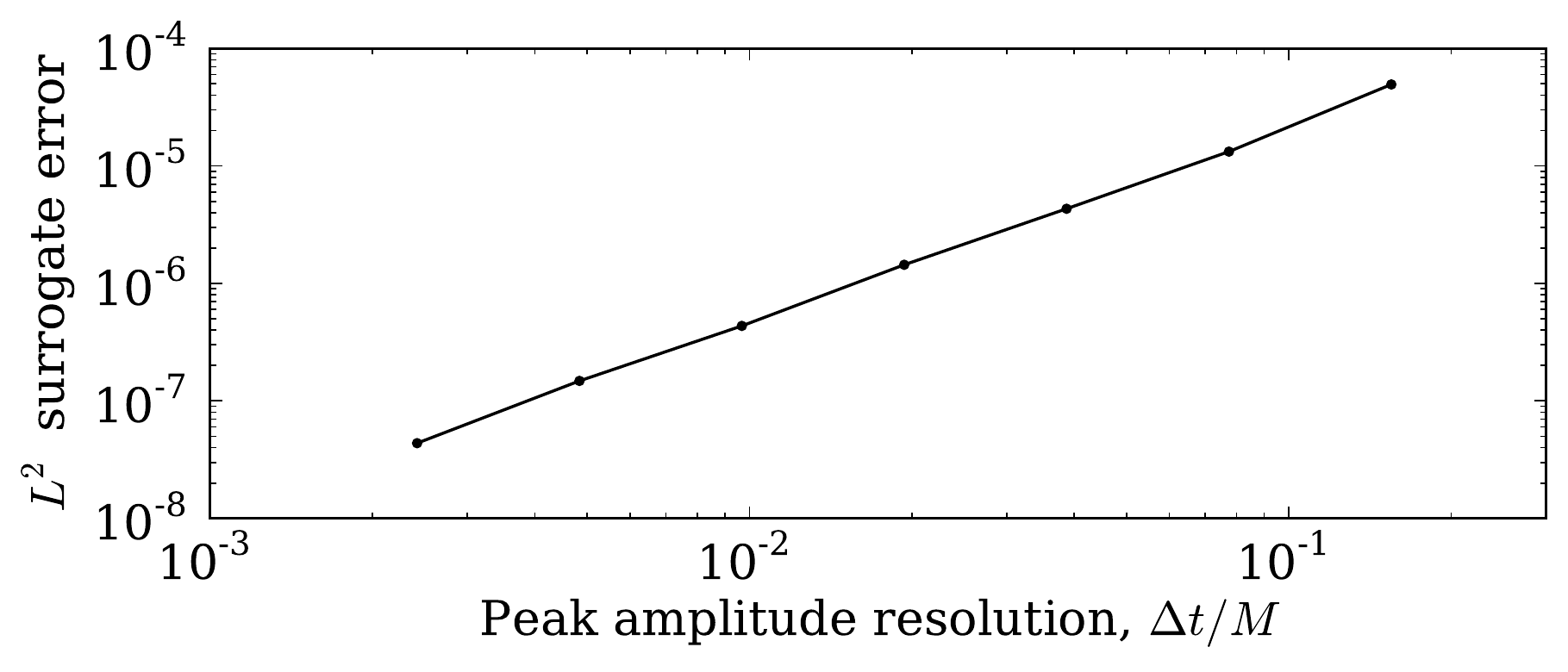}
\caption{The dependence of the error (\ref{eq:l2_surrogate_error_d}) 
when using the surrogate to model an EOB waveform (with $q=1.068$ from Fig.~\ref{fig:EOB_ExampleSurrogateError}) as a function of the resolution (i.e., time steps) of the peak amplitude. The trend is linear in $\Delta t/ M$. The resolution leads to an uncertainty in estimating the peak amplitudes and thus into aligning the waveforms. This is the dominant source of error in the surrogate model that translates directly into errors in the fits of the last offline step for building the surrogate.}
\label{fig:EOB_ExampleFinerRes}
\end{figure}

Aligning the waveforms in the manner discussed above is expected to depend smoothly on the mass ratio $q$ since the time of maximum amplitude, measured from the start of an orbital evolution with a fixed $f_\mathrm{min}$, is expected to depend smoothly on $q$. In practice, waveforms are only known at time intervals $\Delta t$ so that each waveform's peak time is determined within $\Delta t$. Consequently, aligning discrete waveforms introduces some degree of  ``non-smoothness.'' We initially found the surrogate's error to be dominated by this effect. To overcome this difficulty, we generated each waveform on a temporal grid of spacing $\Delta t_\mathrm{fine}$, which allowed the time of maximum amplitude to be resolved within $\Delta t_\mathrm{fine}$. Next, we downsampled each waveform to a sampling rate of interest, say $2,\!048$Hz, such that the peak was located on the downsampled grid. Once a downsampled waveform is generated, neither building nor evaluating the surrogate carries a cost that depends on $\Delta t_\mathrm{fine}$ in any way. Such observations are not unique to surrogate modeling. Indeed, other applications that align waveforms, especially those that need (or expect) some degree of smoothness with parametric dependence, will encounter similar issues. Figure \ref{fig:EOB_ExampleFinerRes} shows how the surrogate error in (\ref{eq:l2_surrogate_error_d}) for our EOB example changes with the grid spacing $\Delta t_{\rm fine}$.
In this paper, the smooth parameter dependence and the aligning of the waveforms at the peak amplitude combine to give fast convergence of the surrogate model to the fiducial one.

\section{Other approaches for waveform prediction} \label{sec:donot}

In this paper we provided a three-step solution for quickly and accurately predicting gravitational waveforms within any given physical model. Here, we discuss a few other approaches that could have been taken instead, which include: i) interpolating the projection coefficients $\{c_i(\mlam)\}_{i=1}^m$, defined from (\ref{eq:projcoeff}), in $\mlam$, ii) interpolating the (complex) waveforms $\{h(T_i; \mlam) \}_{i=1}^m$ at each empirical time, and iii) fitting the amplitudes and phases at {\it all} times. The first approach is an alternative to the empirical interpolation in Step 2 and the fitting in Step 3, the second approach is an alternative to the fitting in Step 3 (Sec.~\ref{sec:LS}), and the third approach is an alternative to empirical interpolation in Step 2 (Sec.~\ref{sec:EIM}). We consider these in turn.

\begin{figure}[ht]
\includegraphics[width=0.98\linewidth]{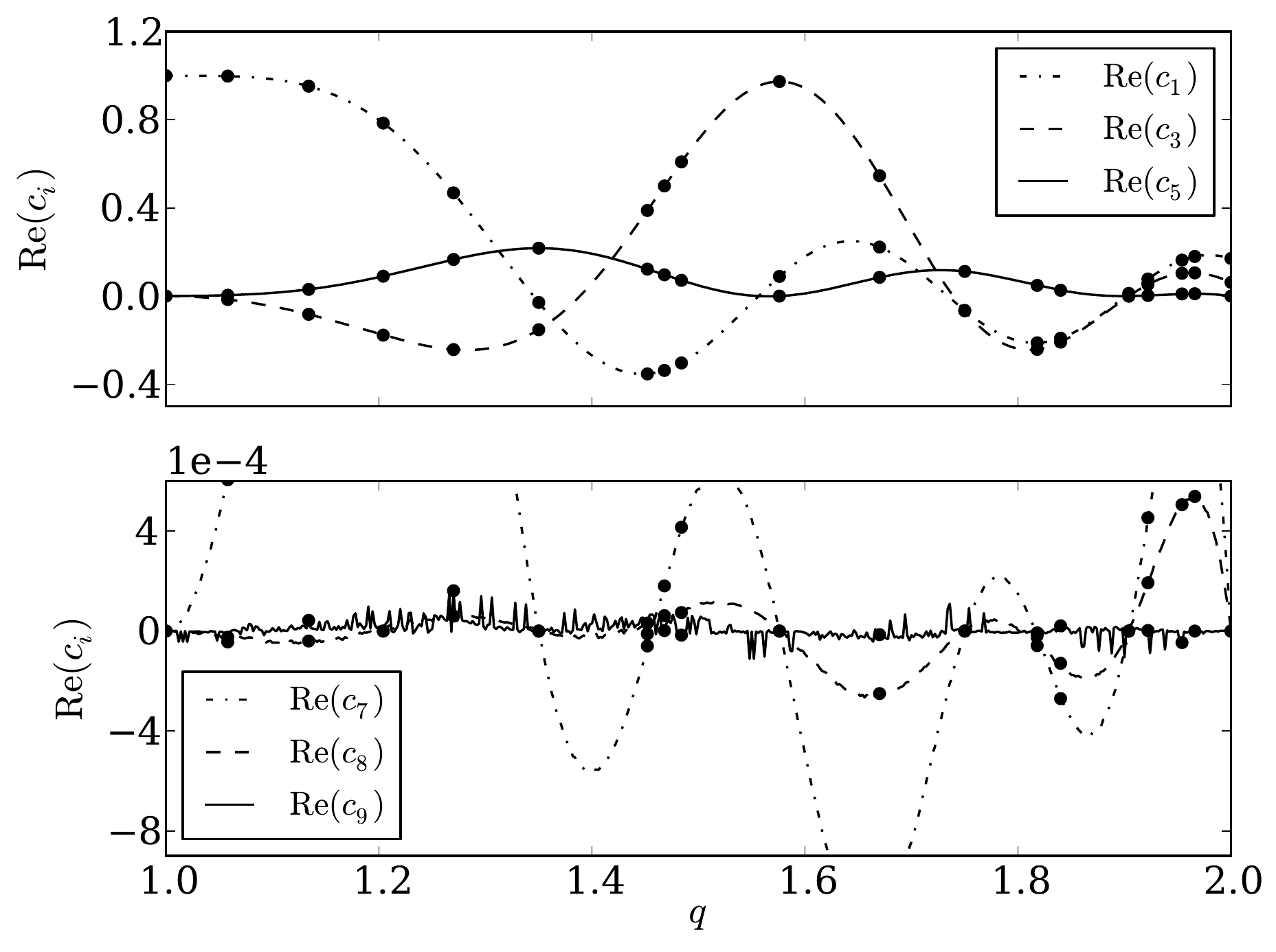}
\caption{The curves depict a variety of projection coefficients $c_i(q)$ along with the greedy data as a function of mass ratio for our EOB example case introduced in Sec.~\ref{sec:SurrogateModels}. Only a representative few curves are shown. The top panel shows the kind of structure that the coefficients have, thereby 
preventing accurate global (polynomial) fits without additional data points. The bottom panel shows the transition in the behavior of these functions with mass ratio from smooth to noisy. 
} 
\label{fig:EOB_FirstProjection}
\end{figure}

The first alternative is to build an interpolating (e.g., Chebyshev) grid in $\mlam$ for each $c_i (\mlam)$. 
This approach was carried out in Refs.~\cite{Cannon:2012gq,PhysRevD.85.081504,brown_sc_2013_13} for inspirals (in the stationary phase approximation in the frequency domain) and phenomenological waveforms for large chirp masses~\footnote{While closed-form expressions provide a useful test bed for experimentation there is less benefits to be gained by interpolating these waveform models over large regions of parameter space. Instead, rapid overlap computations can be directly achieved through specialized reduced order quadrature formula~\cite{antil2012two,Canizares:2013ywa} or (when using a fixed set of templates) by pre-computing the project coefficients as is currently done in the Low Latency Online Inspiral Detection pipeline~\cite{Cannon:2011vi}.}. Problems with such an approach include: i) waveforms from binaries with many GW cycles require increasingly dense interpolation grids \cite{brown_sc_2013_13}, ii) the number of grid points scales exponentially with the number of parameter dimensions, iii) standard grid-based interpolation is not hierarchical and dictates sampling locations at predetermined points that are not tailored to the waveform family of interest, and iv) grids that are essential to resolving one projection coefficient may not be useful for resolving another projection coefficient. Finally, the projection coefficients can be poorly behaved as functions of $\mlam$ because they stem from nontrivial overlaps between waveforms and the basis. This is confirmed using our nominal EOB example, as shown in Fig.~\ref{fig:EOB_FirstProjection} where some of the coefficients become noisy, and also in Ref.~\cite{brown_sc_2013_13}. Furthermore, using only the $m$ greedy points, which comprise an unstructured and under-sampled grid for this problem, exacerbates many of the aforementioned problems.

\begin{figure}[ht]
\includegraphics[width=0.98\linewidth]{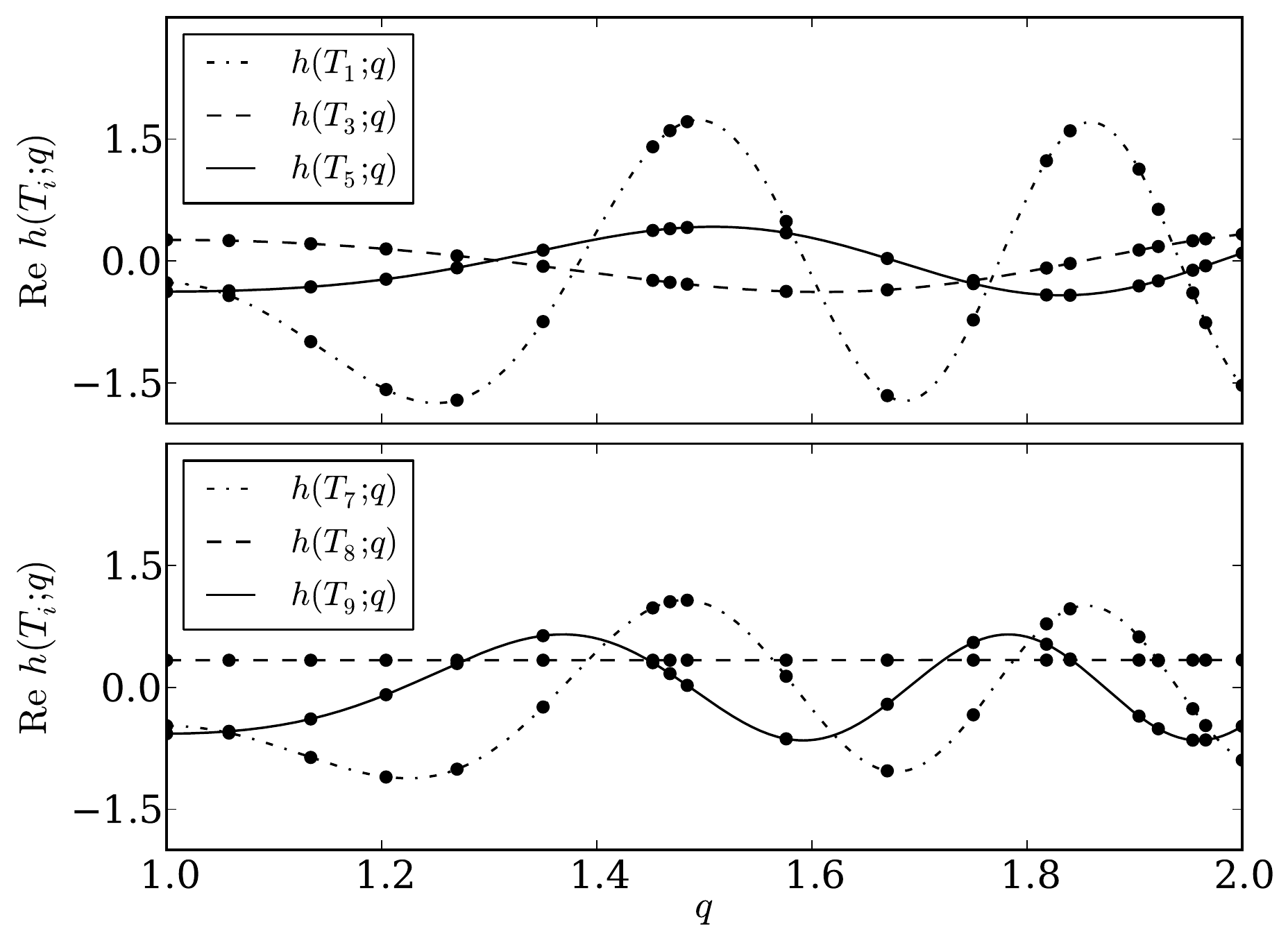}
\caption{The curves depict the values of the real parts of the waveforms along with the greedy data as a function of the mass ratio for our EOB example case. Only curves at a few representative empirical times are shown. While there is less structure here than appears in the coefficients shown in Fig.~\ref{fig:EOB_FirstProjection}, the majority of functions still require additional sampling to be accurately resolved by global polynomial fits.}
\label{fig:EOB_ExampleDEIMPoint3}
\end{figure}

The second alternative is to interpolate in $\mlam$ the complex waveforms at each empirical time. This approach has the same problems as interpolating the projection coefficients discussed above. Figure (\ref{fig:EOB_ExampleDEIMPoint3}) shows the structure of the waveforms as a function of mass ratio at several empirical times in our nominal EOB example. 

The third alternative is to perform fits for the waveform amplitude and phase at \emph{all} time samples instead of the ones dictated by the EIM. It is instructive to compare the operation counts for the online evaluation between this all-times fitting alternative and our EIM-based method. If $c_{\rm fit}$ is the operation count of the fitting functions at each time, taken to be constant for simplicity, then the dominant operation count is $2c_{\rm fit}L$ for the all-times fitting and $2m(L+c_{\rm fit})$ using EIM and fitting at each empirical time. Therefore, making the reasonable assumption that $m \ll L$, the EIM-based approach is more efficient whenever
\be
c_{\rm fit} \gtrsim m \, . \label{eq:cost_comparison} 
\ee

In one parameter dimension, the standard way of evaluating a polynomial fit of degree $n$ is through Horner's algorithm \cite{horner}, which has an optimal operation count of $2n$. It would then seem to follow from (\ref{eq:cost_comparison}) that the online evaluation cost of the EIM-based approach is comparable to fitting at all $L$ times. However, operation counts can be misleading as they do not take into consideration other aspects of an algorithm's implementation that are also relevant for the total execution time. We conducted numerical experiments with our nominal EOB example and found that, for our particular implementation, fitting at all $L \approx 10,\!000$ samples is between $20$ and $1,\!000$ times slower. These timing experiments depend sensitively on both the number of surrogate basis/nodes as well as using ``vectorized'' for-loops. 
Therefore, the actual online evaluation cost in the examples considered in this paper are consistently an order of magnitude or more faster than what a naive operation count would suggest. 

The operation count for evaluating polynomials grows with the dimensionality. 
While the most efficient scheme for evaluating multivariate polynomials is not presently known \cite{Lodha97aunified}, it is 
an active area of research. In general, (\ref{eq:cost_comparison}) is easily met by surrogate models in higher parameter dimensions and we expect the EIM-based surrogate approach to be more efficient than one based on fitting at all output times. 
In addition, the cost to construct $L$ 
separate fits in higher parameter dimensions could make this offline step prohibitively expensive.

\bibliographystyle{physrev}
\bibliography{references}

\end{document}